\pdfoutput=1
\documentclass[keywords]{iucr}

\journalcode{J}
\usepackage{soul}
\usepackage{amsmath}
\usepackage{amssymb}
\usepackage{booktabs}
\usepackage{graphicx}
\usepackage{xcolor}
\usepackage{tikz}
\usepackage[table]{xcolor}
\definecolor{glintgreen}{RGB}{225,242,225}
\usetikzlibrary{arrows.meta,positioning,shapes.geometric,calc}
\usepackage{url}

\makeatletter
\renewcommand\paragraph{\@startsection{paragraph}{4}{\z@}%
  {0.8ex plus 0.3ex minus 0.2ex}%
  {-0.5em}%
  {\normalfont\normalsize\bfseries}}
\makeatother

\begin{document}

\renewcommand{\thefigure}{\arabic{figure}}

\newcommand{\SItitle}[1]{\vskip 16pt plus8pt minus2pt{\centering\rmfamily\large\bfseries #1\par}\vskip6pt\noindent}
\newcommand{\FIXME}[1]{\hl{[FIXME: #1]}}
\newcommand{\SUBMITMODE}{}
\definecolor{revblue}{RGB}{0,90,200}
\makeatletter
\@ifundefined{SUBMITMODE}{%
  \newcommand{\rev}[1]{{\color{revblue}#1}}%
}{%
  \newcommand{\rev}[1]{#1}%
}
\makeatother
\definecolor{movedorange}{RGB}{200,95,0}
\makeatletter
\@ifundefined{SUBMITMODE}{%
  \newcommand{\movedon}{\color{movedorange}}%
  \newcommand{\movedoff}{\normalcolor}%
  \newcommand{\revon}{\color{revblue}}%
  \newcommand{\revoff}{\normalcolor}%
}{%
  \newcommand{\movedon}{}%
  \newcommand{\movedoff}{}%
  \newcommand{\revon}{}%
  \newcommand{\revoff}{}%
}
\makeatother
\newcommand{\SIsec}[1]{\vskip 12pt plus8pt minus2pt{\centering\rmfamily\normalsize\bfseries #1\par}\vskip3pt\noindent}

\title{\rev{Real-time} blind indexing by cross-frame consensus}
\shorttitle{Blind consensus indexing}


\cauthor[a]{Stefano}{Marchesini}{smarches@stanford.edu}{}
{\renewcommand{\corrmark}{}\cauthor[a]{Yuan}{Ni}{yn754@stanford.edu}{}}
\aff[a]{SLAC National Accelerator Laboratory, Menlo Park, California 94025, \country{USA}}
\shortauthor{Marchesini and Ni}

\maketitle

\begin{synopsis}
GLINT combines blind unit-cell determination by cross-frame consensus with GPU-accelerated known-cell registration, indexing sparse serial diffraction patterns \rev{at rates compatible with the real-time data pipelines of high-repetition-rate facilities}, without requiring a unit cell in advance.   
\end{synopsis}

\begin{abstract}
Classical crystallography determines orientation and unit-cell geometry from a rotation series collected from a single specimen, whereas serial crystallography acquires one still exposure from each of many randomly oriented microcrystals and must reconstruct a complete dataset by pooling partial measurements across the ensemble. In serial experiments, each frame typically contains only a subset of reflections, and frames with few reliable peaks \rev{(the small-$N$ regime)} are especially difficult to index \rev{blindly}, because multiple incorrect lattices or orientations can explain sparse observations.

The approach implemented in GLINT addresses this small-N regime by treating the dataset, rather than the individual frame, as the unit of inference. Candidate solutions are generated for many frames, weak but recurrent lattice hypotheses are pooled across
the ensemble to identify a shared unit cell, and each frame is then registered against that consensus cell through a known-cell orientation search with lower dimensionality than blind indexing. %

In benchmark tests, GLINT matched \rev{the strongest blind indexer in the comparison} on indexing yield \rev{at roughly $40\times$ lower per-frame blind-indexing latency against XGANDALF's fastest configuration}, and batched known-cell registration ran entirely on one GPU, allowing cell discovery on the fly during serial data collection. On experimental serial data, frames indexed blind by GLINT merged to crystallographic quality ($CC^*  = 0.90$), while the method rejected non-crystal images rather than forcing unsupported indexing assignments. These results demonstrate that indexing performance can be evaluated not only by indexing yield and throughput, but also by the quality of the resulting merged crystallographic data. 

\end{abstract}

\section{Introduction} \vspace{0pt}

Serial crystallography (SX) determines crystal structures by combining diffraction measurements from a large number of crystals in random orientations. In conventional rotation crystallography, diffraction data are collected from a single crystal over a series of controlled orientations. In SX, by contrast, each microcrystal typically contributes a single still exposure. At an X-ray free-electron laser (XFEL), such as the Linac Coherent Light Source (LCLS)~\cite{emma2010lcls,liang2015cxi,sierra2019mfx}, this exposure is produced by one X-ray pulse. A complete crystallographic data set must therefore be assembled from measurements of many nominally identical crystals, each observed at an unknown orientation \cite{chapman2011,crystfel}. 

Indexing is the step that assigns Miller indices to the observed Bragg reflections by determining the crystal orientation and, \rev{in blind indexing}, when the lattice is not known in advance, the unit-cell parameters required to place reflections on a common reciprocal lattice. Indexed reflections can then be integrated, scaled and merged across patterns to obtain structure-factor data for structure determination \cite{Gevorkov2019XGANDALF,crystfel,toro}. 

\rev{Real-time indexing has been an important goal at X-ray facilities because it can convert streaming diffraction images into crystallographic information during beamtime, improve experimental decision-making/steering, and reduce dependence on prior knowledge of unit-cell parameters. This goal is part of the
broader effort to move analysis into the experimental data stream, including the
long-standing proposal that detector-side analysis could assess data quality
during acquisition and inform experimental steering \cite{camera2015}\rev{, and streaming
analysis systems that close the loop in practice: at synchrotron beamlines, tomographic
reconstruction has been run on remote computing during acquisition with a control feedback loop
deciding when an experiment has acquired enough data \cite{bicer2017}. Serial crystallography has
since acquired real-time processing pipelines of its own, which index at facility rates but against
lattice parameters supplied in advance, with on-the-fly cell determination identified as future
work \cite{White2025Real-time}}.}

\begin{figure}
\centering
\includegraphics[width=0.9\textwidth]{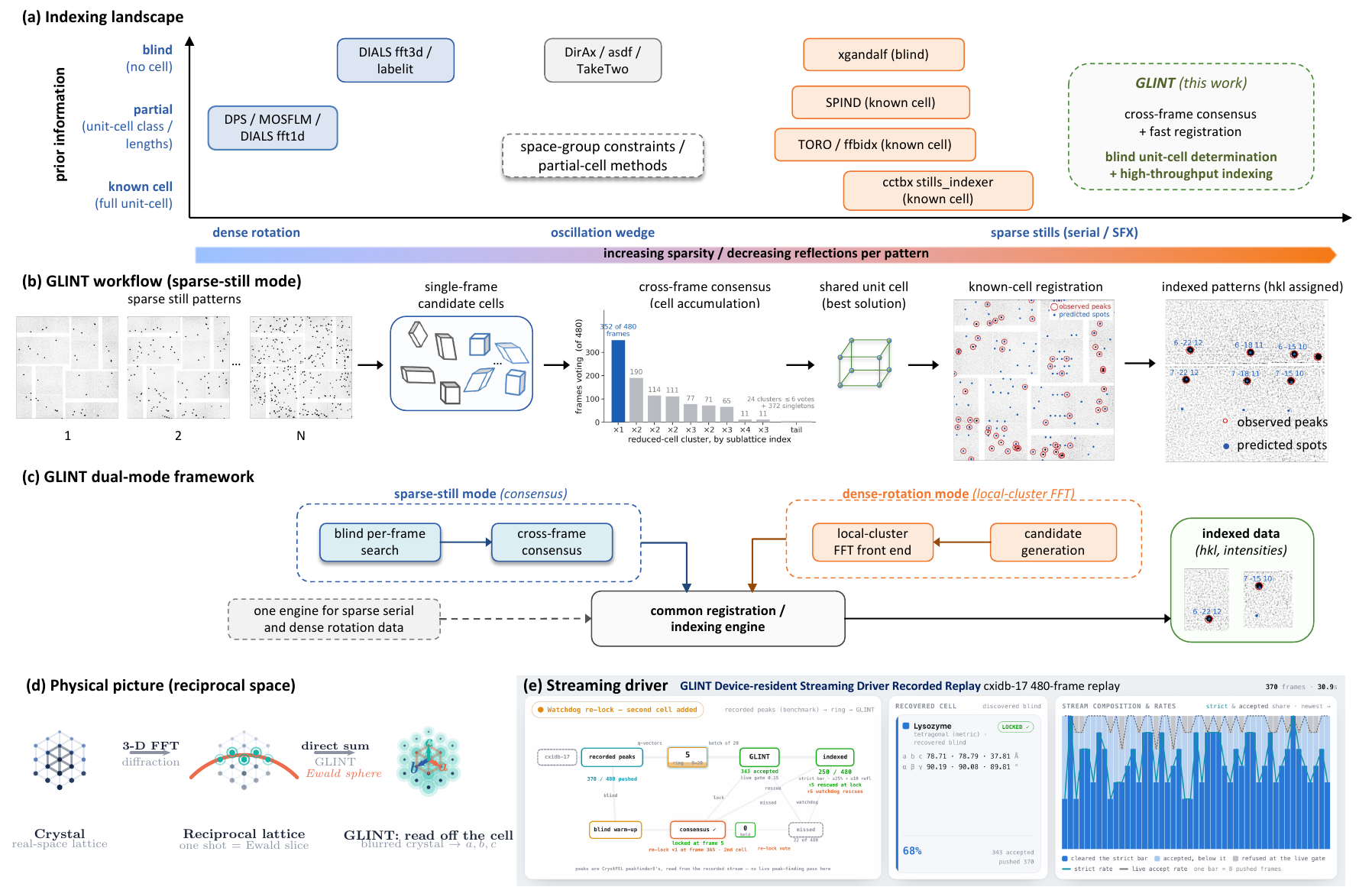}
\caption{Overview of GLINT. (\emph{a}) Indexing landscape: methods placed by reciprocal-space sampling regime (horizontal, dense rotation data to sparse serial stills) and by the unit-cell prior required (vertical: blind, partial, known). Software shown: DPS \cite{dps}, MOSFLM \cite{mosflm}, DIALS \texttt{fft1d} and \texttt{fft3d} \cite{dials}, \texttt{labelit} \cite{labelit}, DirAx \cite{dirax}, CrystFEL \texttt{asdf} \cite{crystfel,crystfel2}, TakeTwo \cite{taketwo}, XGANDALF \cite{Gevorkov2019XGANDALF}, SPIND \cite{Li2019SPIND}, TORO \cite{toro}, the fast-feedback indexer \cite{ffbidx} and the cctbx stills indexer \cite{cctbx,hattne2014}; methods exploiting space-group or partial-cell constraints are marked generically. Quantitative comparisons are given in Tables~\ref{tab:summary} and~\ref{tab:cells}. (\emph{b}) \rev{GLINT} Sparse-still workflow: \rev{each sparse still pattern is first indexed blindly to obtain several
candidate unit cells, which are then pooled across frames to identify the
recurrent cell using the consensus criteria. Once the shared unit cell is
identified, it is used for known-cell registration of unresolved patterns
(open circles, observed peaks; filled circles, predicted reflections).} %
(\emph{c}) \rev{GLINT operates in two sampling regimes with a }dual-mode framework: the sparse-still consensus path \rev{(b)} and the dense-rotation local-cluster FFT front end. \rev{The latter is used for densely sampled reciprocal-space data. Both modes feed a common registration and indexing engine.} 
\rev{(\emph{d}) A reciprocal-space view of GLINT. The Fourier relation connects the two operating modes through the relationship between the reciprocal-space peak distribution and real-space lattice periodicity. A still exposure samples a thin section of the reciprocal lattice near the Ewald
sphere. GLINT recovers the cell vectors (i.e., $\mathbf a,\mathbf b,\mathbf c$) by a grid-free direct sum for sparse
stills, and by local-cluster FFTs for densely sampled data.}
\rev{(\emph{e}) Real-time streaming interface of GLINT during a recorded replay, using the 480-frame cxidb-17 data set, showing online cell discovery, lock status,
and indexing readout (Fig.~\ref{fig:streaming}, Sec.~\ref{sec:streamresults}).
}
}
\label{fig:family}
\end{figure}

\rev{Meeting this goal places several requirements on an indexing method.} Modern serial beamlines are able to generate large streams of still images at high repetition rates \cite{wiedorn2018}\rev{, with detectors being developed for 35~kHz frame rates \cite{king2023}}. \rev{At these rates the raw stream cannot all be retained, so facility data systems reduce it online \cite{thayer2016,thayer2025,galchenkova2024}: Bragg-peak lists are extracted and what is kept is compressed, from ROI-based schemes such as ROIBIN-SZ \cite{underwood2023} to learned representations \cite{ni2026}, and this reduction already runs in detector-side hardware at kilohertz rates \cite{leonarski2023}. An online indexer must therefore consume reduced, streamed data rather than archived frames.}
Furthermore, \rev{indexing methods useful for online use must operate} on continuous data streams, robust on sparse patterns with throughput compatible with rapid experimental feedback \cite{Kieffer2025Application,Gevorkov2019XGANDALF,White2025Real-time}. Existing methods address different parts of this problem. XGANDALF provides blind or optional-cell indexing with strong performance on few-spot serial patterns \cite{Gevorkov2019XGANDALF}; SPIND also targets sparse patterns but uses prior unit-cell information \cite{Li2019SPIND}. GPU-oriented approaches such as TORO and the fast-feedback indexer achieve high throughput when the unit cell is supplied \cite{toro,ffbidx}. Classical Fourier-, projection- and difference-vector-based methods, including DIALS, MOSFLM, CrystFEL \texttt{asdf} and TakeTwo, span dense rotation and still-data regimes \cite{dials,mosflm,crystfel,taketwo}. \rev{Compressive-sensing formulations of auto-indexing have also been explored for femtosecond nanocrystallography \cite{maia2011}.} Figure~\ref{fig:family}(a) summarizes these complementary operating regimes.

\onecolumn
\twocolumn

We propose a \rev{novel} GPU-accelerated, cross-frame indexing method that \rev{improves} blind unit-cell determination and sparse-pattern robustness with high-throughput known-cell orientation registration. \rev{It is motivated by the observation} that serial patterns have different orientations but, for nominally identical crystals, share the same unit cell.  \rev{The method works by retaining multiple candidate cells from blind per-frame solves and pools
them across frames to identify the recurrent lattice. This consensus step
promotes compatible hypotheses and recovers frames that remain unresolved
from a single pattern, thereby improving blind-indexing accuracy. Once the
shared unit cell is determined, the remaining frames can be indexed by
high-throughput known-cell orientation registration.}

\rev{The proposed blind indexing methodology is implemented as part of the blind-indexing front end in GLINT, a new GPU-accelerated indexing framework.}
GLINT also provides a local-cluster Fourier front end for densely sampled reciprocal-space data and an \rev{end-to-end} device-resident streaming driver for online cell discovery and registration. GLINT features: \newline
\textbf{i)} a cross-frame consensus mechanism for unit-cell determination that breaks
the single-frame accuracy ceiling and rejects non-crystals; \newline \textbf{ii)} a batched GPU indexing engine for high-throughput known-cell registration; \newline \textbf{iii)} a
translation-invariant local-cluster transform for candidate generation from densely sampled reciprocal-space data; \newline \textbf{iv)} a device-resident streaming driver that supports cell discovery, registration, and self-locking readout with real-time operational QA.

On real serial diffraction data, GLINT attains blind indexing accuracy comparable with XGANDALF \rev{\cite{Gevorkov2019XGANDALF}} \rev{while reducing the
per-frame indexing latency by roughly $40\times$\rev{, to $26$~ms from about $1.0$~s for a single XGANDALF instance in its fastest configuration (\texttt{-{}-xgandalf-fast-execution}); against CrystFEL's default XGANDALF settings, which are $11\times$ slower, the ratio is roughly $450\times$ (Sec.~\ref{sec:benchmarks})}.
Latency and throughput are distinct measures. Additional CPU cores increase
XGANDALF throughput, reducing the corresponding advantage to
$\sim\!14\times$ against a fully occupied 32-core socket, but do not reduce
the per-frame latency. For real-time operation,
per-frame latency determines how quickly an indexing result becomes available
for experimental feedback, and is therefore reported separately
(Table~\ref{tab:summary}).} On one A100 GPU, batched known-cell registration requires $0.17$~ms per frame of throughput at batches of 120 frames, corresponding to approximately $5.9$~kHz, and the complete locked streaming pipeline, including peak finding, processes approximately $275$ frames~s$^{-1}$. Concurrent work by Nasser \emph{et al.}~\cite{nasser2025} addresses challenging sparse patterns with a symmetry-aware lattice-decoding objective when approximate cell parameters are available; GLINT instead focuses on blind cell determination through cross-frame inference followed by registration. For dense rotation data, the local-cluster front end is compared with the DPS/labelit projection method \cite{dps,labelit} in Sec.~\ref{sec:benchmarks}.

The remainder of the paper is organized as follows. Section~\ref{sec:notation} defines the indexing problem and notation, Sections~\ref{sec:singleframe} and~\ref{sec:multiframe} introduce the single-frame coincidence limit and the cross-frame consensus mechanism, and Sections~\ref{sec:algorithm} and~\ref{sec:streaming} describe the offline and streaming algorithms. Section~\ref{sec:arch} presents the implementation architecture and Section~\ref{sec:lineage} the candidate-generation front ends. The experimental evaluation then examines the single-frame ceiling, consensus, head-to-head indexing performance, streaming operation and its operational diagnostics, end-to-end crystallographic quality, and robustness to multiple lattices and non-crystal data (Sections~\ref{sec:ceiling}--\ref{sec:multilattice}). Section~\ref{sec:throughput} addresses computational performance, followed by data and code availability and an outlook.

\section{Methods}
\subsection{Problem formulation and notation}\label{sec:notation}\label{sec:indexing_problem}
Using the experimental geometry and wavelength, indexing begins by mapping the positions of Bragg reflections on the detector to three-dimensional reciprocal-space vectors. \rev{Concretely, a peak at laboratory position $\mathbf r$, with the incident beam along $+\hat{\mathbf z}$, is mapped to $\mathbf q=(\mathbf r/|\mathbf r|-\hat{\mathbf z})/\lambda$, so that $\mathbf q$ is in \AA$^{-1}$ without a factor of $2\pi$ and $|\mathbf q|=1/d$.} Consider a diffraction pattern $f$, let
\[
\mathcal{Q}_f
=
\{\mathbf{q}_{fi}\}_{i=1}^{N_f},
\qquad
\mathbf{q}_{fi}\in\mathbb{R}^{3},
\]
denote the set of observed reciprocal-space peaks, where $N_f$ is the
number of detected peaks. We write
$\mathcal{Q}_f^{\ast}\subseteq\mathcal{Q}_f$ the subset generated
by a crystal lattice, and
$\mathcal{Q}_f^{\mathrm{sp}}=\mathcal{Q}_f\setminus
\mathcal{Q}_f^{\ast}$ the remaining peaks, which may arise from
background scattering, detector noise, or peak-finding errors.

Let $\Lambda_f$ denote the reciprocal lattice associated with one
crystal in frame $f$.  For a correctly indexed pattern, there exist real-space lattice basis vectors $\mathbf{a}_f$, $\mathbf{b}_f$ and $\mathbf{c}_f$ such that, for each $\mathbf{q}\in\mathcal{Q}_f^{\ast}$,
\[
\mathbf{q}\cdot\mathbf{a}_f=h,\qquad
\mathbf{q}\cdot\mathbf{b}_f=k,\qquad
\mathbf{q}\cdot\mathbf{c}_f=l,
\]
with $(h,k,l)\in\mathbb{Z}^{3}$. Defining the oriented real-space
basis matrix
$
M_f=
\begin{pmatrix}
\mathbf{a}_f & \mathbf{b}_f & \mathbf{c}_f
\end{pmatrix},
$
the corresponding reciprocal lattice can be written as
\[
\Lambda_f
=
\left\{
\mathbf{q}\in\mathbb{R}^{3}:
M_f^{T}\mathbf{q}\in\mathbb{Z}^{3}
\right\}.
\]
Thus, in the ideal case,
$\mathcal{Q}_f^{\ast}\subseteq\Lambda_f$. For experimental data, this relation is approximate, and the fractional-index residual is
\begin{equation}
r(M_f,\mathbf{q})=
\left\|M_f^{T}\mathbf{q}-\operatorname{round}\!\left(M_f^{T}\mathbf{q}\right)\right\|_{\infty}.
\label{eq:residual}
\end{equation}
Indexing therefore seeks an $\widehat{M}_f$, for which a sufficiently large subset of the observed peaks has small residual.

Define the rotation-invariant unit-cell parameters
\[
\boldsymbol{\theta}_f
=
(a_f,b_f,c_f,\alpha_f,\beta_f,\gamma_f),
\]
where $a_f=\|\mathbf{a}_f\|$, $b_f=\|\mathbf{b}_f\|$ and
$c_f=\|\mathbf{c}_f\|$ are the basis-vector lengths, and
$\alpha_f=\angle(\mathbf{b}_f,\mathbf{c}_f)$,
$\beta_f=\angle(\mathbf{a}_f,\mathbf{c}_f)$ and
$\gamma_f=\angle(\mathbf{a}_f,\mathbf{b}_f)$ are the mutual angles.

In known-cell indexing, $\mathbf{\theta}$ is supplied and only the orientation of $M_f$ is unknown. In blind indexing, both the cell and its orientation must be inferred from the peaks.

\begin{figure}
\centering
\resizebox{\columnwidth}{!}{%
\begin{tikzpicture}[
  font=\footnotesize,
  >={Latex[length=2mm]},
  box/.style={draw, rounded corners=3pt, align=center, inner sep=4pt, line width=0.9pt},
  concept/.style={box, text width=48mm, fill=gray!8, draw=gray!60},
  stage/.style={box, text width=74mm, fill=orange!6, draw=orange!75!black},
  rescue/.style={box, text width=74mm, fill=teal!5, draw=teal!60!black},
  keep/.style={box, text width=30mm, fill=green!7, draw=green!55!black},
  dec/.style={box, text width=48mm, fill=gray!2, draw=gray!60},
  ar/.style={->, thick, gray!45!black},
]
\node[stage] (blind) {\textbf{Pass 1 --- blind, every frame} \emph{(combinatorial)}\\
  find + assemble three free vectors from candidates $\rightarrow N_{\mathrm{best}}{=}3$ cells per frame};
\node[concept, above=7mm of blind, xshift=-28mm, fill=teal!5, draw=teal!60!black] (ck) {\textbf{Known cell $=$ registration}\\
  3 rotation DOF; lengths \& angles \emph{given}};
\node[stage, below=5mm of blind] (cons) {\textbf{Consensus} over the pooled hypotheses\\
  the true cell recurs across frames, aliases scatter $\rightarrow$ consensus cell $M_c$};
\node[dec, below=5mm of cons] (dec) {is this frame consistent with $M_c$?};
\node[rescue, below=11mm of dec] (resc) {\textbf{Pass 2 --- rescue \emph{by registration}} against $M_c$\\
  rotate the known basis $M_c$ (3 DOF) until it fits};
\node[keep, right=8mm of dec] (keep) {\textbf{yes:} keep top-1, or promote an $N$-best cell (\emph{free})};
\node[align=center, font=\itshape, below=6mm of resc] (out) {indexed frame $\rightarrow I/\sigma$};
\draw[ar] (blind) -- (cons);
\draw[ar] (cons) -- (dec);
\draw[ar] (dec) -- node[midway,left,font=\scriptsize]{no} (resc);
\draw[ar] (dec) -- (keep);
\draw[ar] (resc) -- (out);
\draw[ar] (keep.south) |- (out.east);
\draw[dotted, thick, gray!70] ([xshift=-14mm]ck.south) to[out=270, in=180, looseness=0.9] (resc.west);   %
\end{tikzpicture}%
}
\vspace{-2mm}
\caption{\textbf{Blind indexing and known-cell registration in GLINT.} The blind pass estimates three unconstrained real-space lattice vectors and retains $N_{\mathrm{best}}$ candidate cells per frame. Cross-frame consensus identifies the recurrent unit cell. Frames already consistent with that cell are retained, whereas unresolved frames are re-indexed by known-cell orientation registration. The streaming driver in Fig.~\ref{fig:streaming} uses the same two operations under a different schedule.}
\label{fig:twopaths}
\end{figure}

\subsection{Consensus breaks the single-frame accuracy ceiling}
\subsubsection{The single-frame problem and accuracy criteria.}\label{sec:singleframe}
For a single pattern $f$, blind indexing estimates an oriented basis $\widehat{M}_f$ using only $\mathcal{Q}_f$. Let $\widehat{\boldsymbol{\theta}}_f$ be the corresponding unit-cell parameters. When a reference lattice $\boldsymbol{\theta}^{\star}$ is available for evaluation, the single-frame cell-identification accuracy over $F$ patterns is
\[
A_{\mathrm{SF}}=
\frac{1}{F}\sum_{f=1}^{F}
\mathbf{1}\!\left[\widehat{\boldsymbol{\theta}}_f\sim\boldsymbol{\theta}^{\star}\right],
\]
where $\sim$ denotes equivalence under the reduced-cell matching criterion associated with eq.~\ref{eq:residual}. %

For a candidate basis $M$, we denote the number of indexed peaks by
$n_f(M)$ and the corresponding peak coverage by
\[
c_f(M)=\frac{n_f(M)}{N_f}.
\]
An indexing acceptance rate is hence the fraction of frames satisfying a
specified gate on lattice correctness, $n_f$, and/or $c_f$. 

Further define inlier count as
\[
s_f(M)
=
\sum_{\mathbf{q}\in\mathcal{Q}_f}
\mathbf{1}
\!\left[
r(M,\mathbf{q})\leq\tau
\right],
\]
where $\tau$ is the indexing tolerance.

A sparse still provides only limited
evidence with which to distinguish the correct lattice from competing
candidates. Consequently, several incompatible bases can explain
comparable subsets of the same observed peaks. More specifically, in a blind search, the true lattice competes with the largest score among many incompatible candidates. If the null scores for incompatible candidates have mean $\mu_{0,f}$ and standard deviation $\sigma_{0,f}$, to leading order, a search over $K$ approximately independent candidates has an extreme-value location of order
\[
s_{\mathrm{floor},f}\simeq\mu_{0,f}+\sigma_{0,f}\sqrt{2\ln K}.
\]
The correct lattice is reliably identifiable only when its score is separated from this look-elsewhere effect.

The same competition can be seen by decomposing the inlier count into crystal and accidental contributions. Let
\[
S_f=|\mathcal{Q}_f^{\ast}|,
\qquad
B_f=|\mathcal{Q}_f^{\mathrm{sp}}|
\]
denote the numbers of crystal-generated and spurious peaks,
respectively. If a fraction $\varepsilon$ of the crystal-generated
peaks fails the inlier test even for the correct candidate, and a
spurious peak matches a reciprocal-lattice node with probability
$p_0$, the expected score of the correct solution is approximately
\begin{equation}
s_{\star}
\approx
S_f(1-\varepsilon)+B_f p_0.
\end{equation}
Note that wrong orientations retain only the accidental component. Searching over $K$ such candidates raises the strongest null
score above the single-candidate background by an amount of order
$\sigma_0\sqrt{2\ln K}$. Hence, successful single-frame indexing requires, schematically,
\begin{equation}
S(1-\varepsilon)\gtrsim\sigma_0\sqrt{2\ln K}.
\end{equation}
We refer to the saturation that occurs when the true-lattice evidence is no longer separated from the strongest accidental coincidences as the \emph{single-frame coincidence ceiling}. Note that the ceiling is not a universal numerical constant: it depends on the cell, the number and distribution of peaks, contamination and the indexing tolerance. Empirical evidence of the ceiling can be found in Sec.~\ref{sec:ceiling}.

\subsubsection{Multi-frame consensus.}\label{sec:multiframe}
Serial data provide information that is absent from an isolated frame. Although crystal orientations vary, nominally identical crystals share the same unit cell. \rev{As implemented in GLINT,} it retains the top $N_{\mathrm{best}}$ distinct cell hypotheses from each blind solve,
\[
\Theta_f=\{\boldsymbol{\theta}_{f1},\ldots,\boldsymbol{\theta}_{fN_{\mathrm{best}}}\},
\]
with $N_{\mathrm{best}}=3$ by default, and reduces them to a rotation-independent cell representation and pools them across frames. More specifically,
\[
\mathcal{P}
=
\bigcup_{f=1}^{F}
\left\{
\mathcal{R}(\boldsymbol{\theta}_{fn})
\right\}_{n=1}^{N_{\mathrm{best}}},
\] where $\mathcal{R}$ denotes Buerger-reduction. A physical lattice accumulates support because it recurs across independently oriented patterns, whereas frame-specific aliases and accidental cells tend to disperse. We call selection of the recurrent lattice from this pooled set \emph{cross-frame consensus}.

If a correct-cell hypothesis is recovered from a fraction $p_{\mathrm{cell}}$ of $F$ frames, its expected support grows as $Fp_{\mathrm{cell}}$. For incoherent spurious hypotheses with local density $\rho_{\mathrm{sp}}$, the background fluctuation grows only as $\sqrt{F\rho_{\mathrm{sp}}}$. The resulting separation therefore scales as
\begin{equation}
\mathrm{SNR}(F)\sim
\frac{Fp_{\mathrm{cell}}}{\sqrt{F\rho_{\mathrm{sp}}}}
\propto\sqrt{F}.
\end{equation}
This scaling explains why pooling independent frames can resolve a lattice that is not the top-ranked solution in every individual pattern. Retaining several hypotheses per frame is important in the sparse regime because the true cell may be reachable but demoted by a frame-specific alias. The empirical coherent and incoherent scaling tests are presented in Sec.~\ref{sec:consensus}.

\subsection{\rev{The GLINT algorithm: per-frame engine and cross-frame consensus}}\label{sec:algorithm}
GLINT operates at two levels (Fig.~\ref{fig:twopaths}). \rev{Inspired by the discussion in Sec.~\ref{sec:multiframe}}, the per-frame indexing engine works by producing the $N$-best cell hypotheses for each
pattern: it evaluates a grid-free direct-sum lattice objective over a Fibonacci-sphere \rev{$\times$} length-shell seed bank, refines every seed in that bank by momentum ascent, then scores and de-duplicates the converged maxima, and assembles bases by triplet enumeration; because every stage is batched tensor arithmetic, all frames and hypotheses are processed concurrently on one GPU, which \rev{enables} retaining $N$-best hypotheses affordable. The cross-frame procedure described here operates on those
hypotheses to determine a consensus unit cell and to recover unresolved
frames. GLINT takes reciprocal-space peak lists from a set of frames and returns either a consensus cell together with per-frame indexing solutions or an explicit refusal if no recurrent lattice satisfies the acceptance criteria. The offline procedure is:
\begin{enumerate}\setlength{\itemsep}{1pt}
\item \textbf{$N$-best blind solve.} Each voting frame is indexed blindly and its top $N$ distinct cell hypotheses are retained ($N=3$ by default).
\item \textbf{Reduce and pool.} Each candidate cell is Buerger-reduced and the hypotheses from all voting frames are collected into one pool.
\item \label{alg:cell_grouping} \textbf{Group by lattice.} Reduced cells are grouped using axis-length, angle-cosine and volume tolerances of $5\%$, $0.06$ and $10\%$, respectively. The batch grouping is seeded in densest-neighbourhood order.
\item \label{alg:consensus_gates} \textbf{Apply consensus gates.} The leading group must contain at least three votes, represent at least $2\%$ of the pool, and exceed the runner-up support by a factor of at least $1.5$. Otherwise no consensus cell is returned.
\item \textbf{Reject derivative aliases (optional).} The leading cell may be compared with its index-$\leq2$ derivative lattices using the coverage--occupancy score; a derivative lattice that provides the tighter systematic explanation causes the candidate to be rejected. This test is disabled at the shipped defaults and was not enabled for any result reported here.
\item \textbf{Report the consensus cell.} The representative cell $M_c$ is the member of the accepted group closest to its median fingerprint.
\item \textbf{Assign frames.} A frame is retained if its top blind solution is consistent with $M_c$. Otherwise the best consensus-consistent retained hypothesis is promoted; frames still unresolved are re-indexed by known-cell registration against $M_c$.
\end{enumerate}

Deriving a representative cell from many indexed patterns is established practice in serial crystallography, for example through inspection of cell-parameter distributions followed by re-indexing with a fixed cell \cite{crystfel,crystfel2}. GLINT incorporates this operation directly into the indexing loop and uses the inferred cell to recover individual frames rather than treating cell determination as a separate manual preprocessing step. \rev{Implementation details are discussed in
Sec.~\ref{sec:arch}.}

\subsection{Streaming: indexing as a live readout}
\label{sec:streaming}
The offline procedure above assumes that a collection of frames is
available before the consensus cell is formed. For continuous
acquisition \rev{and real-time applications}, GLINT reorganizes the same blind-indexing, consensus and
known-cell-registration operations into a stateful
\emph{streaming driver} that updates its cell estimate as frames
arrive (Fig.~\ref{fig:streaming}). 

\begin{figure}
\resizebox{\columnwidth}{!}{%
\begin{tikzpicture}[
  >={Stealth[length=1.6mm]}, font=\scriptsize,
  bx/.style={rounded corners=1pt, draw=black!55, align=center, inner sep=2.2pt,
             minimum height=6.5mm, minimum width=22mm},
  ig/.style={bx, fill=black!4,   draw=black!45},
  fa/.style={bx, fill=teal!9,    draw=teal!60},
  sl/.style={bx, fill=orange!11, draw=orange!65},
  ot/.style={bx, fill=green!9,   draw=green!55},
  ar/.style={->, draw=black!60, shorten >=1pt, shorten <=1pt},
  bulk/.style={ar, line width=1.0pt, draw=teal!70},
  da/.style={ar, dashed, draw=red!30!black!50},
  lb/.style={font=\tiny, text=black!65, inner sep=1.5pt},
  hd/.style={font=\tiny\bfseries, inner sep=1pt}]

\def\cA{0mm}    \def\cB{40mm}   \def\cC{82mm}        %

\def\rA{2mm}
\def\rB{-12mm}
\def\rC{-24mm}
\def\rD{-36mm}
\def\rE{-48mm}
\def\rF{-60mm}

\node[ig] (frames) at (\cA,\rA)  {detector frames\\[-1pt]{\tiny all raw data}};
\node[ig] (pf)     at (\cA,\rB)  {peakfinder $\to q$\\[-1pt]{\tiny hits}};
\node[ig] (veto)   at (\cB,\rB)  {veto\\[-1pt]{\tiny blanks}};
\node[fa] (kc)     at (\cA,\rC)  {known-cell index \textbf{(D5)}\\[-1pt]{\tiny batch 120 $\cdot$ $0.17$\,ms/f}\\[-1pt]{\tiny steady-state throughput}};
\node[fa] (buf)    at (\cB,\rC)  {priority buffer\\[-1pt]{\tiny newest-first $\cdot$ bounded}};
\node[sl] (blind)  at (\cC,\rC)  {blind index \textbf{(D1$\cdot$D6)}\\[-1pt]{\tiny $\times N$ $\cdot$ parallel}};
\node[sl] (cons)   at (\cC,\rD)  {consensus \textbf{(D2)}\\[-1pt]{\tiny support $\geq3$ $\cdot$ seq.\ stop}};
\node[sl] (alias)  at (\cC,\rE)  {alias gate \textbf{(D3)}\\[-1pt]{\tiny refuse super-cell}};
\node[ot] (cell)   at (\cC,\rF)  {cell parameters\\[-1pt]{\tiny active cells $\cdot$ clen}};
\node[ot] (out)    at (\cA,\rE)  {live output\\[-1pt]{\tiny indexed $\cdot$ merge}};

\node[hd, teal!55!black]   at ($(kc.north)+(0,3.4mm)$)    {FAST $\cdot$ known cell};
\node[hd, orange!60!black] at ($(blind.north)+(0,3.4mm)$) {SLOW $\cdot$ blind, parallel};

\draw[ar]   (frames) -- (pf);
\draw[ar]   (pf)     -- (veto);
\draw[bulk] (pf)     -- (kc);
\draw[bulk] (kc)     -- node[lb, left, pos=0.72]{indexed} (out);
\draw[ar] ([yshift=1.6mm]buf.east)  -- node[lb, above]{fan-out} ([yshift=1.6mm]blind.west);
\draw[da] ([yshift=-1.6mm]buf.east) -- node[lb, below]{retry}   ([yshift=-1.6mm]blind.west);
\draw[ar]   (blind)  -- node[lb, right]{vote}     (cons);
\draw[ar]   (cons)   -- (alias);
\draw[ar]   (alias)  -- (cell);

\draw[ar] ([yshift=1.6mm]kc.east)   -- node[lb, above]{miss}        ([yshift=1.6mm]buf.west);
\draw[ar] ([yshift=-1.6mm]buf.west) -- node[lb, below]{\textbf{D4}} ([yshift=-1.6mm]kc.east);

\draw[ar] (alias.west) -- ++(-4mm,0) -- ++(0,8mm)
          -| node[lb, above, pos=0.32]{lock $\cdot$ new cell} ([xshift=7mm]kc.south);

\node[lb, anchor=north west, align=left] at ($(out.south west)+(0,-5mm)$)
     {\textcolor{teal!70}{\rule[0.35ex]{5mm}{1.0pt}}~indexed bulk\\[1pt]
      \tikz[baseline=-0.5ex]{\draw[dashed, draw=red!30!black!50, line width=0.4pt] (0,0)--(5mm,0);}~\emph{retry}: optional blind retry, proposed, not shipped};
\end{tikzpicture}}

\vspace{-2mm}
\caption{\textbf{Streaming organization of GLINT.} A device-resident known-cell path processes frames against the current locked cell, while unresolved frames can optionally enter a bounded miss buffer and a parallel blind path. The blind path supplies the initial cell through sequential consensus and can optionally re-lock when a new recurring cell is detected. Alias checking is applied before a candidate cell is committed, and buffered frames can be re-indexed after a lock. The bold labels are the six streaming operations defined in the text: blind warm-up and adaptive re-lock (D1, D6), sequential-stop consensus (D2), alias gate (D3), miss-buffer rescue (D4) and steady-state known-cell registration (D5). The dashed \emph{retry} edge is a proposed option and is not part of the shipped driver.}
\label{fig:streaming}
\end{figure}

The streaming driver consists of six operations. During \emph{blind warm-up (D1)}, unresolved frames are blind-indexed and contribute $N$-best hypotheses to the running vote. \emph{Sequential-stop consensus (D2)} accepts a cell when the support and runner-up separation criteria are satisfied. An optional \emph{alias gate (D3)} tests the candidate against low-index derivative lattices before it is committed. After a lock, buffered frames can be re-indexed in a batched \emph{miss-buffer rescue (D4)}, and subsequent frames are processed in \emph{steady state (D5)} by known-cell registration. Finally, an optional \emph{adaptive re-lock (D6)} uses accumulated misses to identify a recurrent cell not explained by the active set. The streaming-specific grouping, buffering and batching choices are described in Sec.~\ref{sec:streamimpl}; their measured lock statistics and live-yield effects are reported in Sec.~\ref{sec:streamresults}; the GPU batching, device-resident execution and kernel optimizations
that determine the steady-state throughput are analyzed in Sec.~\ref{sec:throughput}.

\section{Implementation}
\subsection{Architecture (M1--M6)}
\label{sec:arch}

\rev{The GLINT per-frame engine converts a set of reciprocal-space peaks
$Q_f=\{\mathbf q_{fi}\}_{i=1}^{N_f}$ into one or more candidate oriented
lattices. The sparse blind path consists of six modules:
\newline
\textbf{M1} initializes candidate real-space lattice vectors by sampling directions
on a Fibonacci sphere and lengths over a prescribed interval. 
\newline
\textbf{M2} scores each candidate
vector $\mathbf v$ using the grid-free direct-sum
objective
\begin{equation}
\Phi_f(\mathbf v)
=
\sum_{i=1}^{N_f}
w_{fi}\,
\mathbf 1
\left[
\left|
\mathbf q_{fi}\!\cdot\!\mathbf v
-
\operatorname{round}
(\mathbf q_{fi}\!\cdot\!\mathbf v)
\right|
<\epsilon
\right]
\cos(2\pi\mathbf q_{fi}\!\cdot\!\mathbf v),
\label{eq:directsum}
\end{equation}
with $w_{fi}=1/\|\mathbf q_{fi}\|$. A true real-space lattice vector
places lattice-consistent reflections close to integer values of
$\mathbf q_{fi}\!\cdot\!\mathbf v$, so their contributions add
coherently. 
\newline
\textbf{M3} refines each seed by continuous ascent of
$\Phi_f$, yielding a candidate set of lattice vectors.
\newline
\textbf{M4} then forms candidate oriented bases
$M=[\mathbf a\,\mathbf b\,\mathbf c]$ from linearly independent triplets
of the refined vectors. 
\newline
\textbf{M5} subsequently refines each assembled basis jointly, as the three vectors are obtained
independently.
For a current basis $M^{(t)}$ at step $t$, Miller indices are assigned by
\begin{equation}
\mathbf h_{fi}^{(t)}
=
\operatorname{round}
\!\left[
(M^{(t)})^{T}\mathbf q_{fi}
\right],
\end{equation}
after which $M$ is refitted by least squares using reflections whose
fractional-index residual is below the current refinement threshold.
Repeating the assignment and refitting steps brings the three independently
estimated vectors into a common lattice basis.
\newline
\textbf{M6} ranks the refined cells using both lattice residual and peak coverage,
so that a candidate explaining only a small subset of reflections cannot
win solely through a small residual, and reduces the selected cells to a
common primitive representation. 

As shown in Fig.~\ref{fig:pipeline}, the blind path runs M1--M6 and retains
the $N_{\rm best}$ distinct cell hypotheses used by the cross-frame
consensus of Sec.~2.3. When a unit cell is already known, GLINT instead
holds its dimensions and angles fixed and searches only over orientation,
while retaining the same GPU-based refinement machinery. The detailed candidate construction, refinement schedule, acceptance
thresholds, lattice reduction are given
in SI Sec.~\ref{sec:si_modules}, GPU implementation are discussed in Sec.~\ref{sec:throughput}.}

\subsection{Candidate generation across sampling regimes}
\label{sec:lineage}

The sparse and dense front ends in Fig.~\ref{fig:pipeline} can be viewed as two representations of the same lattice-periodicity problem. The connection between them can be made explicit by
representing the reciprocal-space peak set as the weighted point
distribution:
\begin{equation}
\rho_f(\mathbf q)
=
\sum_{i=1}^{N_f}
w_{fi}\,
\delta\!\left(
\mathbf q-\mathbf q_{fi}
\right).
\label{eq:peak_distribution}
\end{equation}
Its three-dimensional Fourier transform is
\begin{equation}
F_f(\mathbf x)
=
\int
\rho_f(\mathbf q)
e^{2\pi i\mathbf q\cdot\mathbf x}
\,d\mathbf q
=
\sum_{i=1}^{N_f}
w_{fi}
e^{2\pi i\mathbf q_{fi}\cdot\mathbf x}.
\label{eq:lattice_transform}
\end{equation}

\begin{figure}
\movedon
\centering
\resizebox{\columnwidth}{!}{%
\begin{tikzpicture}[
  node distance=3.6mm and 4mm, >={Stealth[length=1.6mm]}, font=\scriptsize,
  bx/.style={rounded corners=1pt, draw=black!55, align=center, inner sep=2.2pt, minimum height=6mm},
  sp/.style={bx, fill=blue!7,   draw=blue!55},
  dn/.style={bx, fill=teal!9,   draw=teal!60},
  co/.style={bx, fill=orange!11,draw=orange!65},
  ot/.style={bx, fill=green!9,  draw=green!55},
  fb/.style={bx, fill=black!3,  draw=black!35, font=\tiny},
  ar/.style={->, draw=black!60, shorten >=1pt, shorten <=1pt},
  da/.style={ar, dashed, draw=black!45},
  lb/.style={font=\tiny, text=black!65, inner sep=1pt}]

\node[bx] (in) {peaks $+$ geometry};
\node[bx, below=of in] (fe) {front end $\cdot$ mode auto\\[-1pt]{\tiny route by peak density}};
\node[sp, below left=5mm and 1mm of fe]  (sparse) {sparse $\cdot$ SFX stills\\[-1pt]{\tiny M1 FSS seeds $\to$ M2 score}};
\node[dn, below right=5mm and 1mm of fe] (dense)  {dense $\cdot$ many peaks\\[-1pt]{\tiny local-cluster 3D-FFT seeds}};
\node[co, below=13mm of fe] (core) {M3--M6 core $\cdot$ GPU\\[-1pt]{\tiny refine $\cdot$ assemble $\cdot$ anneal $\cdot$ score}};
\node[fb, right=3mm of dense.east, anchor=west, text width=21mm]
     (fbnote) {fallback\\ starved $\to$ FSS\\ big cell $\to$ adaptive FOV};
\node[bx, below=of core] (cells) {per-frame cell(s)};
\node[sp, below=of cells] (cons) {cross-frame consensus $+$ known-cell index\\[-1pt]{\tiny sparse; residual frames $\to$ cascade}};
\node[fb, left=3mm of cons.west, anchor=east, text width=20mm]
     (casc) {cascade fallback\\ unindexed $\to$ GPU\\ ffbidx / XGANDALF-compatible fallback};
\node[bx, below=of cons] (stream) {CrystFEL \texttt{.stream}};
\node[ot, below left=5mm and -6mm of stream]  (merge) {\texttt{-{}-fromfile} $\to$ CrystFEL\\ refine $\to$ \texttt{partialator}\\[-1pt]{\tiny best merge (refined, tPc)}};
\node[ot, below right=5mm and -6mm of stream] (integ) {\texttt{-{}-integrate} $\to$ $I/\sigma$\\[-1pt]{\tiny quick QC (unrefined)}};

\draw[ar] (in) -- (fe);
\draw[ar] (fe) -- (sparse); \draw[ar] (fe) -- (dense);
\draw[ar] (sparse) -- (core); \draw[ar] (dense) -- (core);
\draw[da] (fbnote.south) |- ([yshift=1mm]core.east);
\draw[ar] (core) -- (cells);
\draw[ar] (cells) -- (cons);
\draw[da] (cons.west) -- (casc.east);
\draw[da] (casc.south) |- ([yshift=2mm]stream.west) node[lb, pos=0.75, above] {residual};
\draw[ar] (cons) -- (stream);
\draw[da] (dense.east) -- ++(2mm,0) |- (stream.east) node[lb, pos=0.25, right] {dense: self-indexes};
\draw[ar] (stream) -- (merge); \draw[ar] (stream) -- (integ);
\end{tikzpicture}}
\vspace{-2mm}
\caption{\textbf{GLINT architecture.} \rev{GLINT operates in two front-ends, the Fibonacci-sphere/direct-sum front end and the 3D-FFT front end for Sparse stills and densely sampled rotation data respectively.} Both routes feed the common M3--M6 refinement, assembly, annealing and scoring stages. Sparse patterns subsequently use cross-frame consensus and known-cell rescue; dense rotation clouds are indexed directly. GLINT can emit a CrystFEL stream for downstream refinement and merging or integrate predicted reflections internally.}
\label{fig:pipeline}
\end{figure}

Note that the real part of
$F_f(\mathbf x)$ is equivalent to Eq.~\eqref{eq:direct_sum} of M2 with
the proximity window removed. The two front ends therefore probe the same underlying lattice
periodicity by different numerical routes. For sparse still patterns,
the M1--M3 path evaluates this periodicity directly at continuous
candidate real-space coordinates, without constructing a global
three-dimensional Fourier grid. For dense rotation data, GLINT instead
uses a local-cluster FFT front end. Because translating a reciprocal-space
peak distribution changes the phase of its Fourier transform but not its
magnitude, small groups of peaks can be recentered and transformed
independently on finely sampled local grids. The Fourier grid is enlarged
when the observed reflection density indicates that longer real-space
lattice vectors, and hence a larger-volume cell, must be represented.
Candidate lattice vectors obtained from the local transforms are subsequently
passed to the common M3--M6.

This Fourier viewpoint (Fig.~\ref{fig:family}(d)) connects to established candidate-generation methods. The autocorrelation
of $\rho_f$, or equivalently the distribution of pairwise peak
differences, has Fourier transform $|F_f(\mathbf x)|^2$ by the
Wiener--Khinchin relation, which connects this formulation to
difference-vector approaches such as DirAx \cite{dirax}. Likewise,
projection-based indexing methods such as DPS \cite{dps} search
one-dimensional Fourier sections of the same three-dimensional
transform, as related by the projection-slice theorem. XGANDALF instead
evaluates the corresponding direct sum at candidate real-space vectors
without first discretizing the transform on a grid
\cite{Gevorkov2019XGANDALF}.

In addition to the two front ends in GLINT, we evaluated
alternative candidate-generation representations, including a global
gridded 3D FFT, pair-difference/autocorrelation constructions, NUFFT
and projection-based searches. These comparisons motivated the
sampling-regime-dependent implementation choices described above.
Detailed algorithms, parameter settings and comparisons across sparse,
intermediate and dense reciprocal-space sampling are given in
Supporting Information Sec.~S3; their effect on indexing performance is
examined in Sec.~\ref{sec:benchmarks}.

\subsection{Streaming}\label{sec:streamimpl}
 The streaming driver implements the six operations
D1--D6 introduced in Sec.~\ref{sec:streaming} as a scheduling layer around
the same blind and known-cell indexing engines used offline.

\paragraph{Acceptance criteria.}
The streaming implementation uses the same reduced-cell equivalence
relation (Sec.~\ref{sec:algorithm}) and the same blind and known-cell indexing engines as the offline procedure (Sec.~\ref{sec:arch}). Its two streaming-specific acceptance criteria are
defined below.

For the D2 sequential-stop vote, let $v_1(t)$ and $v_2(t)$ denote the
supports of the leading and second-ranked reduced-cell groups after
$t$ voting frames, and define the leading support fraction
\[
p_1(t)=\frac{v_1(t)}{t}.
\]
The required leader--runner-up gap is
\begin{equation}
g(t)=
\begin{cases}
2, & p_1(t)\geq 0.30,\\
3, & 0.12\leq p_1(t)<0.30,\\
4, & p_1(t)<0.12.
\end{cases}
\label{eq:stream_gap}
\end{equation}
D2 accepts a cell when
\begin{equation}
\begin{gathered}
v_1(t)\geq 3,
\qquad
v_1(t)-v_2(t)\geq g(t),
\\
\text{and, once }n_{\mathrm{pool}}(t)\geq 72\text{:}
\qquad
v_1(t)\geq 0.02\,n_{\mathrm{pool}}(t),
\qquad
v_1(t)\geq 1.5\,v_2(t),
\end{gathered}
\label{eq:stream_lock}
\end{equation}
where $n_{\mathrm{pool}}(t)$ is the number of candidate cells pooled by
frame $t$.
Thus, the running vote uses the same lattice groups as the offline
consensus but, while the hypothesis pool is small and the final pool
size is not yet known, defers the batch pool-share and
leader-to-runner-up ratio gates in favour of an absolute support gap;
once the pooled hypothesis count reaches $72$, those two gates are
re-applied at their batch values of $2\%$ and $1.5\times$. The gap is
widened when the leading cell recurs only slowly, requiring stronger
separation before committing to a lock.

D1 supplies blind $N$-best hypotheses to the running
vote, D2 applies Eqs.~\eqref{eq:stream_gap}--\eqref{eq:stream_lock}, and
D3, when enabled, applies the derivative-lattice alias test before a
candidate is committed. After a lock, D4 can reprocess buffered frames against the
recovered cell (opt-in; the miss buffer is empty by default), D5 performs
steady-state known-cell registration using
Eq.~\eqref{eq:stream_frame_accept}, and D6 uses accumulated misses to
test for a recurrent cell not explained by the active set.

A frame is accepted in the steady state when enough of its peaks are
near-integer under the registered matrix, in both count and fraction:
\begin{equation}
\bigl|\left\{\, i : r(M_f,\mathbf{q}_{fi}) < \tau \,\right\}\bigr|
\;\geq\;
\max\!\left(n_{\min},\; \nu N_f\right),
\label{eq:stream_frame_accept}
\end{equation}
where $M_f$ is the consensus cell registered in the orientation of
frame $f$ and $r$ is the fractional-index residual of
Eq.~\eqref{eq:residual}, with tolerance $\tau=0.15$, minimum inlier
count $n_{\min}=6$ and minimum inlier fraction $\nu=0.15$ by default.
The same test gates every acceptance decision in the streaming path,
including the warm-up and miss-buffer rescues; frames that fail it are
counted and dropped, or retained for later rescue when the opt-in miss
buffer is enabled.

The offline consensus, by contrast, forms groups in
densest-neighbourhood order and reports the member of the winning group
closest to its median reduced-cell fingerprint. The running vote groups
hypotheses greedily in arrival order and retains the representative of
the winning online group. A hypothesis that falls within tolerance of
several existing groups is treated as evidence that arrival order split
one lattice; those groups are coalesced and the arriving hypothesis
becomes the representative of the merged group, since it is the only
member known to lie within tolerance of all of them. The two procedures can therefore report
slightly different representative cells within the same
tolerance-defined lattice group. On the data sets evaluated here the two
orders select the same lattice; where the detector geometry is poorly
refined, arrival-order grouping can instead select a different, typically
derivative, lattice, which is why the running vote re-applies the batch
share and lead gates once the hypothesis pool is large enough.

Incoming frames enter through a common \texttt{push} interface and can
originate from either a live detector or replayed files. \rev{Two
implementation choices are worth stating, since neither follows from the
operations themselves.} In D1 a buffered warm-up can rank frames by detected
Bragg-peak count and blind-index only the highest-ranked subset before the
remaining frames are processed. Frames not explained by the active cell can be
retained in a bounded miss buffer, which is an opt-in feature and is disabled at
the shipped defaults; the D6 fan-out over accumulated misses then recovers
same-cell failures or supplies hypotheses for a candidate re-lock. \rev{Implementation details are given in Supporting
Information Sec.~\ref{sec:si_impl}; the shipped default parameters and the
streaming stress tests in Sec.~\ref{sec:si_streamstress}.}

The live path applies Eq.~\eqref{eq:stream_frame_accept} at its
permissive defaults ($n_{\min}=6$, $\nu=0.15$), whereas offline
evaluation scores against a stricter bar ($n_{\min}=10$, $\nu=0.25$).
After a cell is locked, D5 does not perform a blind retry on every
frame. The driver can nevertheless optionally emit low-confidence
frames together with their observed peaks --- a per-frame coverage
threshold flags them and a stream-output mode attaches the observed peak
list; the two options are coupled --- so an optional blind retry can be applied later
without increasing live-path latency. The resulting difference in
per-frame yield is quantified in Sec.~\ref{sec:streamresults}.

\section{Results}
\label{sec:results}

\rev{The experimental data are from the Coherent X-ray Imaging Data Bank (CXIDB), which was established in part to support method development for the large data volumes generated by X-ray free-electron
lasers \cite{cxidb,maia2016}.} The numerical experiments are designed to answer the following questions: the single-frame limit of blind indexing \rev{(Sec.~\ref{sec:ceiling})}, the gain obtained by pooling
independent frames \rev{(Sec.~\ref{sec:consensus})}, \rev{the behaviour of the streaming driver (Sec.~\ref{sec:streamresults}),} and performance comparison across different reciprocal-space
sampling regimes \rev{(Sec.~\ref{sec:benchmarks})}, together with the end-to-end quality of the merged data
\rev{(Sec.~\ref{sec:realdata})} and robustness to multiple lattices and non-crystal images \rev{(Sec.~\ref{sec:multilattice})}. Unless stated otherwise, single-frame cell-identification results in
this subsection use the reduced-cell correctness criterion defined in
Sec.~\ref{sec:singleframe}; stricter coverage-gated values are reported
separately where relevant.
\paragraph{Computation.} Unless noted otherwise, GLINT timings are measured on one NVIDIA A100 GPU; CPU indexers are run on the same 32-core AMD EPYC 7542 host. Detailed data provenance and benchmark configuration are provided in the Supporting Information (S1--S2).

\subsection{Data and evaluation}
\label{sec:data}

The principal \textit{sparse-still} benchmark uses the
cxidb-17 lysozyme data set \cite{boutet2012}, recorded with a CSPAD
detector at LCLS. We use the first 120 deposited hit frames as a
controlled benchmark and a 480-frame extension of the same run for
higher-statistics consensus and streaming tests. The 480-frame set uses
the same \texttt{peakfinder8} peak lists from the CrystFEL stream and the
same strict acceptance gate, with the 120-frame benchmark embedded as its
first 120 frames, on which the head-to-head counts of
Table~\ref{tab:summary} \rev{(GLINT $92/120$, xgandalf $86/120$)} are reproduced exactly. The 120-frame subset
contains 40--554 detected peaks per frame (median 100; interquartile
range 63--175) and is defined solely by deposition order, without
selection based on indexing outcome. Detailed data curation is discussed in SI S1.

Additional \textit{serial datasets} test transfer across proteins, unit cells
and detector technologies. These include cxidb-45 Proteinase K
\cite{cxidb,masuda2017}, cxidb-62 ACG (\emph{Agrocybe cylindracea} galectin)
\cite{yamashita2017},
cxidb-83 beta-lactamase \cite{wiedorn2018}, cxidb-61 POMGnT1
\cite{yamashita2017}, and a Jungfrau-4M lysozyme data set recorded at LCLS-CXI (experiment cxil1015922, run 0033). Two
multi-thousand-frame LCLS runs, mfxl1038923 r0278 and r0058, are used
to test consensus stability and multi-lattice behaviour. The
Proteinase K comparison with DIALS uses the same capped list of 170
high-SNR peaks per frame for both indexers, while the end-to-end GLINT
analysis uses its uncapped peak lists. Detector geometry, wavelength
handling and peak-finding details \rev{for the cxidb-17, cxidb-45, Jungfrau-4M and mfxl1038923 sets} are given in Supporting Information
Sec.~S1.

\textit{Synthetic data} are used where the relevant variables cannot be controlled independently in experimental measurements. These tests
include Ewald-sliced lysozyme stills with prescribed spurious-peak
load and mosaic broadening, nanoBragg \cite{nanobragg} simulations with known
structure factors, and noise-free full-rotation reciprocal-lattice
sets spanning the seven lattice systems, and the 560-event streaming-QA
sequence of Sec.~\ref{sec:qa}, a planted six-species mixture with
periodic protein-reference and powder-calibrant events on a simulated
low-$q$ panel. They isolate the
single-frame coincidence model, consensus scaling and dense-rotation
candidate generation from detector-specific effects. Simulation
parameters and generation procedures are given in Supporting
Information Secs.~S1\rev{, S6 and S13}.

\subsection{Empirical evidence for the single-frame accuracy ceiling}
\label{sec:ceiling}

The 120-frame cxidb-17 benchmark was used to test whether the
single-frame saturation, as defined in Sec.~\ref{sec:singleframe}, reflects
ambiguity in the observed reciprocal-space peaks rather than a
particular optimizer or scoring choice. Blind indexing identifies the
correct lattice on 85 of 120 frames (71\%). Under the stricter
criterion requiring the correct lattice and at least 25\% of the
observed peaks to be indexed, 79 of 120 frames (66\%) are accepted.
Increasing the density of the candidate search, retaining more
candidates, changing the final scorer and introducing alternative
single-frame refinement procedures did not produce a consistent
increase beyond this range. The corresponding ablations are summarized
in Supporting Information Table~S3.

\begin{figure}
\includegraphics[width=0.9\columnwidth]{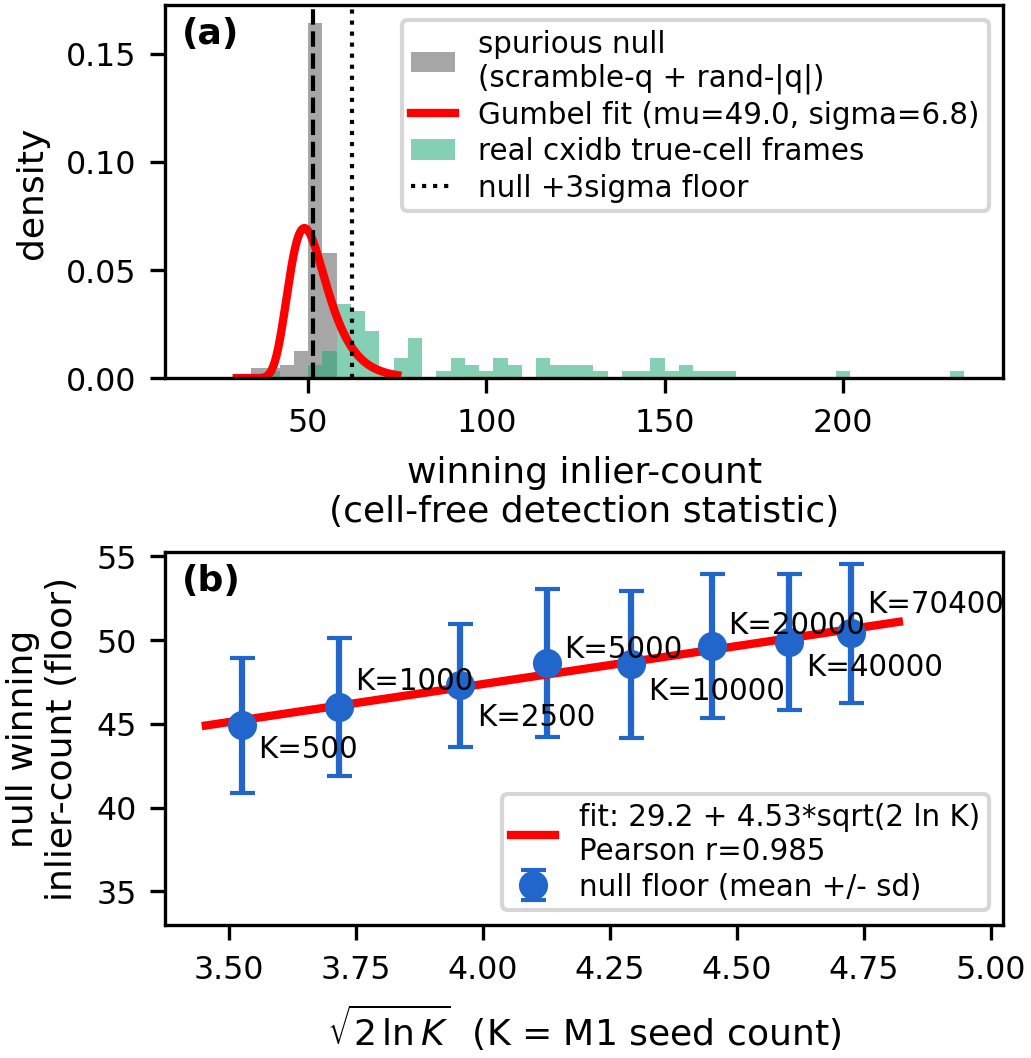}
\caption{\textbf{Extreme-value floor for the single-frame blind search.} (\emph{a}) Distribution of the maximum inlier count for the $K=70{,}400$-seed search applied to spurious-only null frames, in which both the direction and the magnitude of every peak vector are randomized (\texttt{scramble-q} plus \texttt{rand-}$|\mathbf q|$), together with the corresponding distribution for real crystal frames evaluated against the reference cell. The null width includes frame-to-frame variation in peak count and contamination. (\emph{b}) Dependence of the null-floor location on search size $K$; the observed growth is consistent with the $\sqrt{2\ln K}$ look-elsewhere scaling over the tested range ($r=0.985$).}
\label{fig:stat_floor}
\end{figure}

The null experiment in Fig.~\ref{fig:stat_floor} tests the
extreme-value mechanism proposed in Sec.~\ref{sec:singleframe}. For the
$K=70{,}400$ candidate seeds used by the sparse search, the
maximum inlier count $s_f(M)$ on spurious-only
(\texttt{scramble-q}) frames forms a Gumbel-like extreme-value
distribution \cite{gumbel1958}. As the number of tested candidates is increased from
$K=500$ to $70{,}400$, the location of this null floor increases
approximately with the predicted $\sqrt{2\ln K}$ dependence, with a Pearson $r=0.985$
(Fig.~\ref{fig:stat_floor} (b)).

When the real crystal frames are evaluated against the reference cell,
$98\%$ ($78/80$) score above the extreme-value floor: the $\sqrt{2\ln K}$ fit
of Fig.~\ref{fig:stat_floor}(\emph{b}) places that floor at $50.6$ inliers at
$K=70{,}400$, consistent with the directly measured null maximum of
$50.4\pm4.2$. The median separation is 7.5 standard deviations of the
null-maximum distribution (mean 11.4), and $75\%$ ($60/80$) lie more than
$3\sigma$ above that distribution's mean. The frames therefore contain strong
evidence for the physical lattice when that lattice is supplied, even
though the blind search does not always select it. At the stricter criterion
the 41 unaccepted frames divide into 33 generation misses, in which no
retained candidate matches the reference lattice, and 8 selection misses, in
which such a candidate is present but is not chosen ($28\%$ and $7\%$ of the
120 frames). Together with the
growth of the null maximum as the candidate search is enlarged, this
supports the interpretation that accidental lattice coincidences contribute
to single-frame failure alongside incomplete candidate generation, rather
than that local refinement is insufficient.

To distinguish failures of candidate generation from failures of final
candidate selection, we performed an oracle decomposition in which all
retained blind hypotheses were inspected against the reference lattice.
A failure was classified as a \textit{generation miss} when no retained
hypothesis matched the reference cell, and as a \textit{selection miss} when a
matching hypothesis was present but was not chosen by the final scorer.
Under the strict gate, generation misses account for approximately
28\% of frames, whereas final-selection errors account for only
6--8\%. As a complementary control, supplying the reference cell to the
known-cell registration path recovers essentially all frames. Thus the
majority of frames missed by blind indexing are indexable crystal
patterns for which the correct lattice is absent from, or demoted
within, the finite single-frame hypothesis set. These observations
identify incomplete candidate information together with accidental
lattice coincidences as the dominant sources of the observed
single-frame saturation on cxidb-17. The complementary known-cell control is repeated on the denser
cxidb-62 lattice in Sec.~\ref{sec:benchmarks}, where supplying the
reference cell indexes $99.8\%$ of frames; the oracle decomposition
itself was run only on cxidb-17.

\subsection{Multi-frame consensus breaks the ceiling}
\label{sec:consensus}
\subsubsection{Cross-frame consensus breaks the accuracy ceiling.} On the 120-frame
cxidb-17 benchmark, the single-frame blind pass indexes 85/120 frames at
the $s_f(M)\geq10$ reflection criterion (on this set the correct-lattice
and $\geq$10-reflection bars coincide for the single-frame blind pass,
so this is the same 85/120 as Sec.~\ref{sec:ceiling}). After pooling the retained
per-frame hypotheses, determining the consensus cell, and re-indexing
unresolved frames by known-cell registration, the accepted fraction rises
to 115/120. The improvement persists under the stricter criterion requiring
the correct lattice and $c_f(M)\geq0.25$, for which the single-frame rate
increases from 79/120 to 92/120.

As a control, supplying the reference cell directly to the same
registration path gives 91/120 frames at the strict gate, essentially the
same result as the 92/120 obtained using the cell inferred by consensus.
Thus, at the strict gate on this data set, the cross-frame estimate
supplies the cell information needed for subsequent registration without
requiring a unit-cell prior. The comparison is gate-dependent: on a
480-frame extension of the same data the supplied reference cell leads
the consensus estimate at the weaker $\geq$10-reflection bar ($468$
against $458$ frames), so the equivalence is stated per gate rather than
in general.

\subsubsection{Scaling with the number of pooled frames.}
\begin{figure}
\includegraphics[width=0.8\columnwidth]{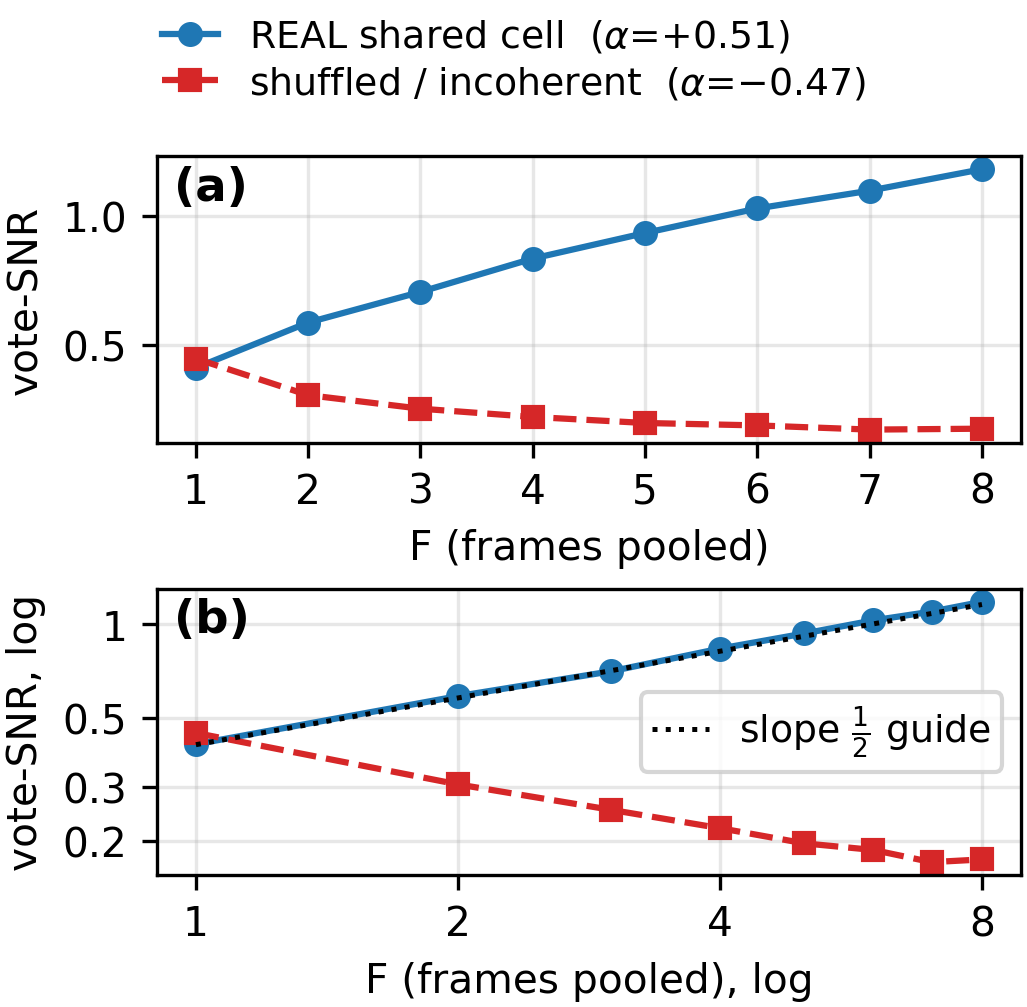}
\caption{
\textbf{Consensus scaling on synthetic Ewald-sliced lysozyme stills}. \rev{With 
spurious load $f=1.0$ (i.e.\ one spurious peak added per lattice
peak; $\sigma=0$)}, both ensembles are simulated; in the legend,
`REAL' labels the genuinely shared-cell arm, not experimental data.
Each of \rev{$F$} pooled frames contributes
its six best blind cell hypotheses --- six rather than the shipped default $N_{\mathrm{best}}{=}3$, to widen the vote histogram --- clustered using the
reduced-cell equivalence of Sec.~\ref{sec:algorithm}; a frame
votes at most once in each cluster. For the shared-cell ensemble,
the vote signal-to-noise is
\rev{$S_{\mathrm{vote}}=H_{\mathrm{true}}/\sqrt{B_{\mathrm{rest}}}$}, where $H_{\mathrm{true}}$
is the true-cell support and \rev{$B_{\mathrm{rest}}$} is the support of all other
clusters. In the shuffled control, peak directions are randomized
isotropically at fixed $|\mathbf q|$, removing the shared lattice
while preserving the radial distribution; because no true cell
exists, $H_{\mathrm{top}}$ and the support of the remaining
clusters are used instead. Power-law fits \rev{$S_{\mathrm{vote}}\propto F^{\alpha}$} give
\rev{$\alpha=0.51$} for the shared-cell ensemble and \rev{$\alpha=-0.47$} for the
shuffled control. The dotted line is an exponent-$1/2$ reference.
}
\label{fig:stat_sqrtn}
\end{figure}

The second test isolates the recurrence mechanism predicted in
Sec.~\ref{sec:multiframe}. Synthetic Ewald-sliced lysozyme patterns were
pooled either with a unit cell shared across independently oriented
frames or under a control in which no cell was shared across the pool.
Passing both ensembles through the same consensus procedure gives
opposite scaling behaviours (Fig.~\ref{fig:stat_sqrtn}). For the
shared-cell ensemble, the fitted dependence has an exponent
$\alpha\simeq0.51$, where $\alpha$ is defined by a power-law fit
$S(F)\propto F^{\alpha}$ (log--log least-squares fit). This is close to the
$F^{1/2}$ scaling predicted in Sec.~\ref{sec:multiframe}. Under the
no-shared-cell control, the corresponding exponent is negative
($\alpha\simeq-0.47$), showing that pooling alone does not increase the statistic when there is no recurrent lattice. 

The number of frames required for a stable consensus depends on the
frequency with which the correct cell appears among the retained
single-frame hypotheses. On the cxidb-17 benchmark, the consensus
statistic approaches its plateau within approximately three to five
frames, and a vote over the first five frames already returns the
full-set cell. We additionally run random-subset tests on two independent long LCLS runs, \textit{mfxl1038923 r0278} (1785 indexed frames) and \textit{r0058} (2319 indexed frames). A five-frame
subset recovered the corresponding all-frame cell in only 36\% and
41\% of draws. Under the reconstructed
protocol, approximately sixteen pooled frames were required to
exceed 90\% recovery on both runs. The random-subset protocol and recovery criterion are given in SI~\ref{sec:si_random}.

\subsubsection{Phase diagram.}
To determine when additional frames can overcome the single-frame
ceiling, we varied two properties of synthetic Ewald-sliced lysozyme
stills independently: the spurious-peak load $f$ (spurious peaks added per lattice peak, so $f{=}0.8$ is $44\%$ of peaks spurious) and the mosaic
broadening $\sigma$. Each $(f,\sigma)$ condition contains 360
simulated frames. We define
$N^{\star}(f,\sigma)
$
as the smallest number of pooled frames for which the correct consensus
cell is recovered in at least 90\% of trials. Thus,
$N^{\star}=1$ denotes a condition already resolved from individual
frames, while $N^{\star}>1$ quantifies the additional cross-frame
information required for reliable cell recovery.

\begin{figure}
\includegraphics[width=0.7\columnwidth]{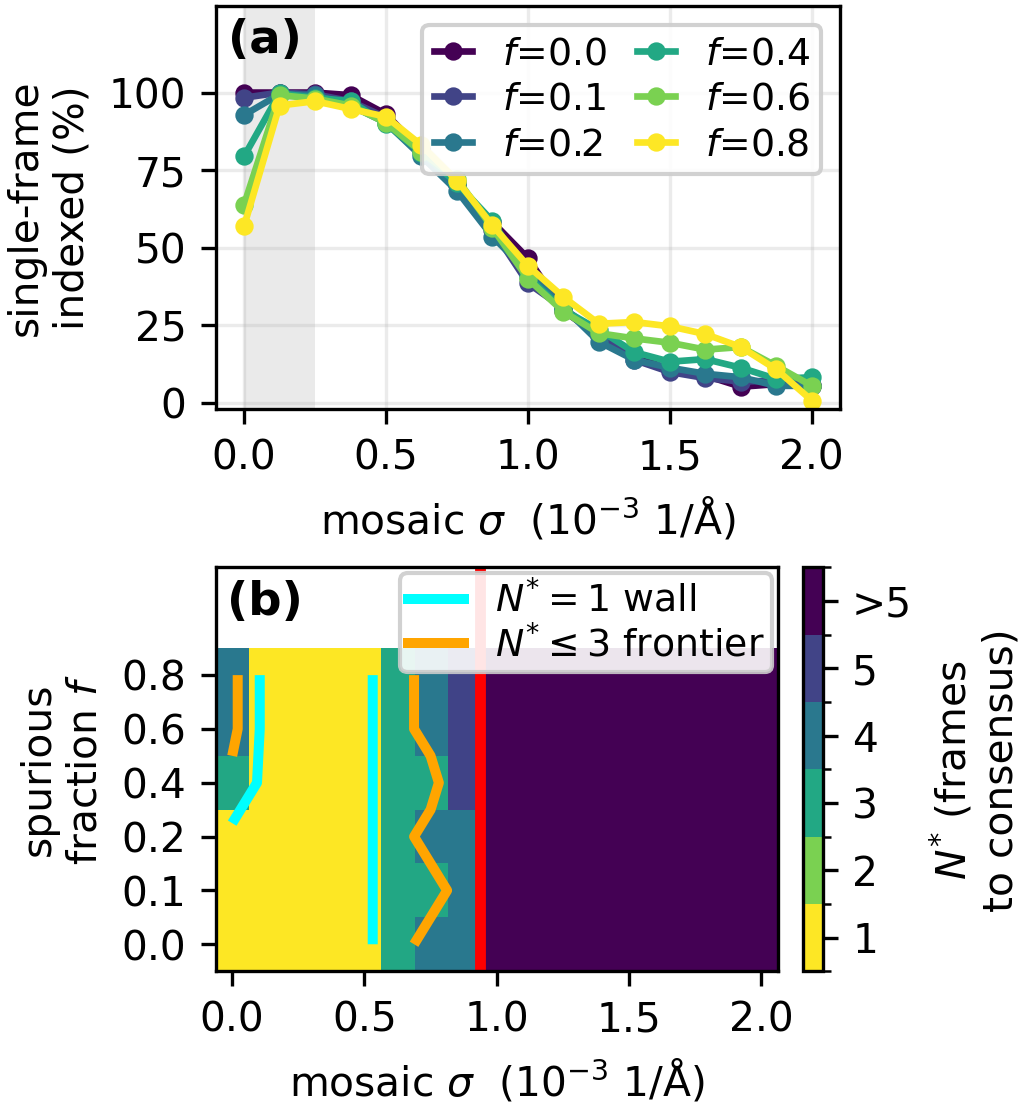}
\caption{\textbf{Single-frame and consensus behavior.}\rev{Over controlled spurious load $f$ (spurious peaks added per lattice peak, so $f{=}0.8$ is $44\%$ of peaks spurious)} and mosaic broadening $\sigma$ for synthetic lysozyme stills (360 frames per condition). (\emph{a}) Single-frame indexed fraction as a function of $\sigma$ for several values of $f$. Moderate broadening initially improves the single-frame rate before mosaic blur exceeds the indexing tolerance; the shaded band marks the $\sigma\leq0.25\times10^{-3}\,\mathrm{\AA}^{-1}$ range over which this improvement holds. (\emph{b}) Minimum number of pooled frames $N^{\star}$ required to reach 90\% consensus. The map separates a single-frame-sufficient region, a consensus-resolvable band, and a generation-limited regime in which additional frames do not restore the missing cell hypotheses. The cyan and orange curves are the $N^{\star}=1$ and $N^{\star}\leq3$ contours, and the vertical red line marks the onset of the generation-limited regime near $\sigma=1.0\times10^{-3}\,\mathrm{\AA}^{-1}$.}
\label{fig:stat_phase}
\end{figure}

The controlled $f\times\sigma$ sweep (Figure~\ref{fig:stat_phase}) reveals three regimes: 

\begin{itemize}
    \item at low ambiguity, the physical cell is identified reliably from a single frame and
$N^{\star}=1$.
    \item In an intermediate band, individual frames frequently
fail, but the correct cell remains present among their retained
hypotheses; pooling two to five frames raises its recurrent support
above the competing frame-specific aliases. This is the
consensus-resolvable regime in which cross-frame inference breaks the
single-frame ceiling.
    \item At sufficiently large mosaic broadening, the correct cell is no longer generated consistently by the per-frame
blind solver, and increasing the pool size therefore cannot restore
reliable consensus. This generation-limited
regime appears near
$\sigma=1.0\times10^{-3}\,\mathrm{\AA}^{-1}$, where even pools larger
than five frames fail to reach the 90\% criterion across the tested
spurious fractions. The outer boundary is controlled primarily by
$\sigma$ in this simulation, whereas changing the spurious fraction
mainly affects the region in which the correct and accidental
hypotheses still compete.
\end{itemize}

\subsection{Streaming Driver}
\label{sec:streamresults}

The streaming driver provides an indexing-rate readout during acquisition. We next test whether the batch consensus gains can extend to sequential constraints of online acquisition, and quantify the per-frame yield associated with the low-latency streaming schedule. 

\subsubsection{Streaming lock and live-path yield.}
Across 400 random arrival orders of the 480-frame cxidb-17 set, the
D2 sequential-stop vote reaches the same lattice as the batch consensus
after a median of six voting frames (mean 7.2; 90th percentile 12).
No conflicting lattice is locked in these reorderings. Thus, for this
data set, the online vote reproduces the batch cell using only a short
prefix of the incoming frames. 

The online schedule yields fewer accepted frames than the full offline
pipeline. At the strict $c_f(M)\geq0.25$
evaluation gate, the live path indexes 67--69\% of the 480-frame set ---
$67\%$ ($323/480$) at the shipped defaults, rising to $69\%$ ($331/480$)
with the two opt-ins, warm-up rescue and adaptive re-lock, enabled ---
compared with $74\%$ ($357/480$) for the offline known-cell reference,
which is handed the textbook cell with consensus skipped (Supporting
Information Sec.~\ref{sec:si_impl}). Cell determination does not account
for the gap: run instead on its own consensus-derived cell, the same
offline pipeline indexes $361/480$ ($75\%$), a tie with the handed cell
on the discordant frames ($16$ against $12$, $p=0.57$). The difference
therefore arises primarily from per-frame scheduling after the lock. In
particular, D5 does not perform a blind retry after every failed
known-cell registration. Applying such a retry only to the frames that
fail the gate closes $98$--$100\%$ of the gap in the synthetic sweep, and
on the real 480-frame set it does more than close it: $365/480$ ($76\%$),
eight frames past the offline pipeline's $357$. On the 120-frame subset
the same retry reaches $88$ against offline's $91$, so the crossover is a
property of run length rather than of the retry itself. That retry is
measured offline on the recorded stream and is not part of the shipped
streaming driver (Fig.~\ref{fig:streaming}). Detailed retry, warm-up and
stress tests are given in Supporting Information
Sec.~\ref{sec:si_streamstress}.

\subsubsection{Operational diagnostics in streaming mode.}
\label{sec:qa}

A drop in indexing yield can result from different failure mode including changes in sample delivery, crystal
quality, beam position or detector geometry. We therefore tested whether including \textit{periodic
reference} measurements can supplement the streaming readout and help attribute a degradation to the indexing path, or to detector geometry. Two complementary reference measurements are used. A protein reference probes the crystallographic indexing path by testing recovery of a known
cell and the corresponding known-cell inlier support. A powder calibrant
probes detector geometry independently, its ring radii was set by the
geometry rather than by which reflections a given orientation excites.
Calibrant selection, radial-integration implementation and
diagnostic thresholds are described in SI
Sec.~S13. 

\rev{The driver additionally reports a running state --- lock status and the consensus support behind it, the composition of the active cell set, alias-gate refusals, multiple-lattice and refused-frame counts, accumulated geometry drift, and recovery activity (e.g., Fig.~\ref{fig:family}(e)) --- all of which are by-products of the indexing process itself. Each responds differently to common failure modes; the individual readouts and the implications of changes in each are discussed in SI Sec.~S13.}

\begin{table}
\caption{
Response of the streaming and reference diagnostics to a planted
2-pixel beam-centre displacement at event 300 of a simulated 560-event
test sequence. The diffraction frames are synthetic; the peak finding,
indexing, consensus and geometry diagnostics are the pipeline's own
outputs (SI Sec.~S13).
}
\label{tab:qa}

\footnotesize
\renewcommand{\arraystretch}{1.15}

\begin{tabular*}{\columnwidth}
{@{\extracolsep{\fill}}lcc@{}}
\toprule
diagnostic & before shift & after shift \\
\midrule
indexed / hit
    & 79.3\%
    & \cellcolor{red!7}\textbf{5.9\%} \\

protein reference passing
    & 2/2
    & \cellcolor{red!7}\textbf{0/3} \\

blind reference-cell recovery
    & 100\%
    & \cellcolor{red!7}\textbf{0\%} \\

calibrant profile correlation
    & 1.0000
    & \cellcolor{red!7}\textbf{0.654} \\
\bottomrule
\end{tabular*}
\end{table}

For a controlled test, we used a simulated 560-event sequence carrying a planted mixture of six protein species, replayed through the real GPU peak-finding and streaming pipeline together with
periodic protein-reference and powder-calibrant measurements. A
2-pixel \textit{beam-centre displacement} was introduced at event 300. The
responses before and after the perturbation are summarized in
Table~\ref{tab:qa}. The planted displacement reduces the indexed fraction from 79.3\% to
5.9\%, so the streaming indexing rate itself rapidly signals that the
data quality has degraded. The calibrant provides complementary
information: its radial-profile correlation decreases from 1.0000 to
0.654, independently indicating that the degradation is consistent with
a change in detector geometry rather than solely with a change in the
sample. Its value in this test is therefore attribution rather than
earlier detection. Sensitivity to smaller geometry perturbations and
detector-specific diagnostic thresholds are examined in Supporting
Information Sec.~S13.

\rev{
\subsubsection{Additional operational considerations.}

Modern facility data systems reduce the stream online, extracting features
such as Bragg peak lists rather than archiving raw frames
\cite{thayer2016,thayer2025}. Hence, at an operational level, real-time blind
indexing must also tolerate such upstream online data reduction, at the rate
peak lists are produced and on hardware co-located with the reduction
pipeline. As an additional robustness test, we examined compatibility with
an LCLS-II-facing, ROIBIN-SZ-style lossy compression pipeline
\cite{underwood2023}. ROIBIN-SZ works by preserving regions around Bragg
peaks while reducing the surrounding detector background, \rev{and learned
representations are increasingly used for the same purpose \cite{ni2026}.} On the 480-frame
cxidb-17 benchmark, the reduction produced no statistically resolved change
in aggregate indexing yield over the tested settings, although individual
frame-level outcomes did change. This test is complementary to the streaming
and crystallographic-validation experiments above and specifically evaluates
the robustness of indexing to this form of upstream data reduction
(Supporting Information Sec.~S17).
}

\subsection{Benchmark comparisons}
\label{sec:benchmarks}

\subsubsection{Sparse stills benchmark on cxidb-17.} On the common 120-frame sparse benchmark using the sparse cxidb-17, GLINT and XGANDALF achieve comparable blind indexing accuracy at the strict acceptance gate (i.e., $c_f(M)\geq0.25$), whereas their computational costs differ by orders of magnitude
(Table~\ref{tab:summary}).

\rev{A single GLINT blind solve requires 26\,ms per frame, compared with about
$1.0$\,s for one XGANDALF instance in its fastest configuration, corresponding to
approximately $40\times$ lower per-frame latency. We quote that configuration
because a comparison should be made against a baseline at its best: CrystFEL's
\texttt{-{}-xgandalf-fast-execution} mode coarsens the sampling pitch and reduces
the gradient-descent iteration count, and on the same host and the same 120
frames it runs $11\times$ faster than the default settings while indexing
$88/120$ rather than $86/120$ at the same gate --- faster \emph{and} marginally
more accurate here, not a speed-for-accuracy trade. Against the default settings
the latency ratio is roughly $450\times$. Throughput, by contrast, depends on the
parallel CPU hardware assigned to XGANDALF: running independent instances across
all 32 cores of the benchmark host reduces the advantage to approximately
$14\times$, and against a current 192--288-core socket it is closer to two.
Additional CPU parallelism narrows the throughput ratio but does not reduce the
latency of an individual XGANDALF solve. Only the latency ratio is invariant to
how much CPU hardware is assigned, and it is the quantity that decides how
quickly an indexing result becomes available for experimental feedback --- which
is what makes real-time operation possible.}
    
On indexing accuracy, GLINT indexes 92/120 frames and XGANDALF 86/120. Importantly, the GLINT single-frame blind front end indexes only
79/120 frames at the same strict gate, below XGANDALF's 86/120.
The increase from 79/120 to 92/120 therefore arises from cross-frame
consensus and subsequent known-cell rescue rather than from a stronger
single-frame blind solve. We extend the test to the 480-frame cxidb-17 dataset (Sec.~\ref{sec:data}), where GLINT-\textcircled{1} indexes 361/480 frames and XGANDALF 350/480; the 57 discordant
frames split 34:23 in GLINT's favour, an exact two-sided McNemar test giving $p=0.18$,
which likewise does not establish a significant accuracy difference. The appropriate conclusion is parity in blind indexing yield under these conditions together with a substantial throughput advantage for the GPU implementation. 

Fourier-, difference-vector- and cell-search back ends perform substantially below the direct-sum methods. As a control, we
applied the same back-projection harness to 60 well-conditioned DIALS
lysozyme frames with substantially richer reciprocal-space sampling.
On this set, \texttt{asdf} and \texttt{mosflm} index 98--100\% of
frames when the cell is supplied and 85--88\% in blind mode. Their
recovery on the richer patterns supports the interpretation that the
separation observed in Table~\ref{tab:summary} arises as
reciprocal-space sampling becomes sparse.

\begin{table}
\caption{\label{tab:summary}Head-to-head indexing on the 120-frame sparse cxidb-17 benchmark. All methods are evaluated on the same reciprocal peak lists under a common strict gate: the recovered lattice must be correct and at least 25\% of observed peaks indexed. The \emph{lattice} column reports the correct-lattice bar alone, without the coverage requirement. GLINT runs on one A100 GPU; CPU methods run on the same 32-core AMD EPYC~7542 host. GLINT-\textcircled{1} denotes blind indexing followed by cross-frame consensus and known-cell rescue. For XGANDALF the table reports single-instance wall time; 32 independent instances on the host give an effective socket-level wall time of approximately 361~ms per frame, so the \rev{$\sim\!40\times$ in per-frame latency against XGANDALF's fastest configuration corresponds to} $\sim\!14\times$ \rev{in throughput} against a fully occupied 32-core socket. \rev{The tabulated XGANDALF ms/frame is CrystFEL's default setting; \texttt{-{}-xgandalf-fast-execution} is $11\times$ faster on the same host and indexes $88/120$ rather than $86/120$, and is the configuration the latency ratio is quoted against (Sec.~\ref{sec:benchmarks}).} $^{\dagger}$\texttt{asdf} and \texttt{mosflm} wall-clock at 32 threads; XGANDALF and DIALS single-instance. $^{\ddagger}$A throughput measurement at batch size $B=120$, not single-frame latency. $^{\S}$A single call; a persistent pipelined configuration sustains approximately 320~frames~s$^{-1}$. $^{\P}$Correct reduced cell, without the $25\%$ coverage requirement, reported where per-frame solutions were available to re-score. This bar is the more stable of the two: the coverage threshold is sharp and two of the 120 frames fall within one indexed peak of it, so the \emph{indexed} counts carry an intrinsic $\pm1$ sensitivity to floating-point ordering that the lattice counts do not.}
\scriptsize
\setlength{\tabcolsep}{1pt}

\begin{tabular}{llrrrr}

indexer & mode & indexed & lattice$^{\P}$ & ms/frame & frames/s \\
\midrule
\rowcolor{glintgreen}
GLINT-\textcircled{1} & blind & \textbf{77\%} (92/120) & \textbf{96\%} (115/120) & \textbf{26} & \textbf{39} \\
GLINT (single-frame blind) & blind & $66\%$ (79/120) & $71\%$ (85/120) & --- & --- \\
xgandalf & blind & \textbf{72\%} (86/120) & $78\%$ (94/120) & 11542 & 0.087 \\
DIALS FFT3D & blind & $27\%$ (32/120) & --- & 2483 & 0.40 \\
DPS (rstbx) & blind & $25\%$ (30/120) & --- & 110 & 9.1 \\
mosflm & blind & $21\%$ & --- & 714$^\dagger$ & 1.4 \\
asdf & blind & $0\%$ (28 spurious) & --- & 143$^\dagger$ & 7.0 \\
xgandalf & known-cell & \textbf{80\%} (96/120) & $92\%$ (111/120) & 9136 & 0.11 \\
GLINT-\textcircled{1} & known-cell & $76\%$ (91/120) & --- & 32 & 31 \\
\rowcolor{glintgreen}
GLINT-\textcircled{1} (batched)$^{\ddagger}$ & known-cell & $76\%$ (91/120) & --- & \textbf{0.17} & \textbf{5900} \\
ffbidx & known-cell & $75\%$ (90/120) & --- & 4.4 & 226$^{\S}$ \\
asdf & known-cell & $35\%$ & --- & 143$^\dagger$ & 7.0 \\
mosflm & known-cell & $27\%$ & --- & 200$^\dagger$ & 5.0 \\

\end{tabular}
\end{table}

\begin{table}
\caption{
Same-input indexing of 907 cxidb-45 Proteinase~K frames using the
same capped list of 170 high-SNR peaks per frame. GLINT consensus
remains blind to the unit cell: the cell is inferred from pooled
frame hypotheses before known-cell rescue.
}
\label{tab:proteinase_dials}

\footnotesize
\renewcommand{\arraystretch}{1.12}

\begin{tabular*}{\columnwidth}{@{\extracolsep{\fill}}llr@{}}
\toprule
method & indexing mode & indexed \\
\midrule
GLINT          & single-frame blind   & 32.0\% \\
DIALS 1D-FFT   & blind                & 61.9\% \\
DIALS grid     & cell supplied        & 49.9\% \\

\addlinespace[2pt]
\rowcolor{green!8}
\textbf{GLINT} & \textbf{consensus + rescue}
               & \textbf{97.8\%} (887/907) \\
\bottomrule
\end{tabular*}
\end{table}

\subsubsection{Cross-frame gain on Proteinase K.}
A same-input comparison with DIALS \cite{dials,hattne2014} on capped Proteinase K peak lists separates the per-frame and cross-frame effects (Table~\ref{tab:proteinase_dials}). This comparison is deliberately chosen so that the per-frame result
does not favour GLINT: DIALS 1D-FFT indexes 61.9\% of frames blindly, compared with 32.0\% for the GLINT single-frame blind front end, showing that the direct-sum front end is not uniformly superior on this large tetragonal cell. Supplying the cell does not rescue the per-frame regime either: the DIALS real-space grid search targeted on the Proteinase~K cell (``DIALS grid'' in Table~\ref{tab:proteinase_dials}) indexes 49.9\%, below DIALS's own blind rate, because these sparse stills are limited by the number of peaks available to fix an orientation rather than by knowledge of the cell; this is one known-cell route within DIALS, not a fully tuned \texttt{dials.stills\_process} configuration. After pooling GLINT hypotheses across frames, however, consensus recovers the cell and known-cell rescue raises the result to 97.8\% on the same capped inputs. The gain therefore arises from the cross-frame architecture rather than from a uniformly stronger per-frame candidate search.

\begin{table}
\caption{\label{tab:cells}Dense-rotation comparison across ten unit cells spanning the seven lattice systems. Identical noise-free full-rotation reciprocal-lattice clouds are indexed by the GLINT local-cluster FFT front end on one A100 GPU and by the DPS/labelit implementation in \texttt{rstbx} on one AMD EPYC~7542 core. Four random orientations are used per cell, so indexing rates are quantized in 25\% increments. The comparison isolates indexing performance from photon statistics and detector effects.}
\footnotesize
\setlength{\tabcolsep}{4pt}
\begin{tabular}{lrrrrrr}
\toprule
 & & \multicolumn{2}{c}{GLINT (GPU)} & \multicolumn{2}{c}{DPS (CPU)} & \\
\cmidrule(lr){3-4}\cmidrule(lr){5-6}
cell (\AA) & rlps & rate & ms & rate & ms & speedup \\
\midrule
lysozyme 79/79/38        & 36.9k & $100\%$ & 44 & $100\%$ & 1136  & $26\times$ \\
Proteinase K 68/68/109   & 79.3k & $\mathbf{100\%}$ & 50 & $75\%$  & 2022  & $40\times$ \\
tetragonal 58/58/130     & 67.8k & $100\%$ & 44 & $100\%$ & 1760  & $40\times$ \\
hexagonal 105/105/75     & 109k  & $100\%$ & 50 & $100\%$ & 2586  & $52\times$ \\
cubic 78/78/78           & 73.5k & $100\%$ & 46 & $100\%$ & 1907  & $42\times$ \\
orthorhombic 60/110/135  & 138k  & $\mathbf{100\%}$ & 52 & $50\%$  & 3168  & $61\times$ \\
triclinic 45/55/65       & 24.4k & $100\%$ & 44 & $100\%$ & 775   & $18\times$ \\
monoclinic 60/70/90      & 56.6k & $100\%$ & 43 & $100\%$ & 1560  & $36\times$ \\
\rowcolor{glintgreen}
tetragonal 140/140/150   & 456k  & $\mathbf{100\%}$ & 73 & $25\%$  & 10652 & $\mathbf{146\times}$ \\
rhombohedral 70/70/70    & 50.4k & $100\%$ & 46 & $100\%$ & 1393  & $30\times$ \\
\bottomrule
\end{tabular}
\end{table}

\subsubsection{Known-cell comparison.}

The known-cell comparison on cxidb-62 tests orientation registration separately from blind cell determination. With the cell supplied, GLINT indexes 99.8\% of 28,835 frames and remains at 98.7\% on the subset with at most 25 reflections (Table~\ref{tab:cxidb62}). The published baseline values for XGANDALF, TORO and the symmetry-aware method of \citeasnoun{nasser2025} are lower\rev{, although acceptance conventions are method-specific and the comparison is cross-study rather than harmonized (Table~\ref{tab:cxidb62})}. Against XGANDALF and TORO the separation widens sharply on the sparsest patterns ($7.2$ and $6.8$ points on all frames, $26.4$ and $24.5$ points at $N_f\leq25$), whereas the symmetry-aware method holds its rate on the sparse subset. This result indicates that the GPU registration path remains robust when few reflections are available, while the blind consensus mechanism addresses the separate problem of obtaining the cell\rev{; on this hexagonal set GLINT's blind single-frame rate is approximately $19\%$ (Supporting Information Table~\ref{tab:realindex})}.

\begin{table}
\caption{
Cell-informed indexing on the cxidb-62 frame set. GLINT values are
measured here; XGANDALF, TORO and \citeasnoun{nasser2025} are the
published rates reported for the same data set. Acceptance conventions
are method-specific, so the table provides cross-study context rather
than the harmonized head-to-head comparison of
Table~\ref{tab:summary}. The sparse subset contains the 3881 frames
with at most 25 observed reflections.
}
\label{tab:cxidb62}

\footnotesize
\renewcommand{\arraystretch}{1.12}

\begin{tabular*}{\columnwidth}{@{\extracolsep{\fill}}lrr@{}}
\toprule
method & all frames & $N_f \leq 25$ \\
\midrule
\rowcolor{glintgreen}
\textbf{GLINT}      & \textbf{99.8\%} & \textbf{98.7\%} \\
Nasser \emph{et al.} & 95.2\% & 95.9\% \\
TORO       & 93.0\% & 74.2\% \\
XGANDALF   & 92.6\% & 72.3\% \\
\bottomrule
\end{tabular*}
\end{table}

\subsubsection{Dense front end comparison.}
The dense-rotation benchmark tests the
\textit{dense · many peaks} front end of GLINT (Fig.~\ref{fig:pipeline}) against the DPS/labelit \cite{dps,labelit}
implementation in rstbx. DPS/labelit is the natural baseline here because it is the one dense-data method that consumes exactly the same input as this front end --- bare reciprocal-lattice-point clouds rather than detector images --- and is callable as a library, so a like-for-like comparison of candidate generation is possible; image-based dense pipelines enter the evaluation instead through the same-input Proteinase~K comparison of Table~\ref{tab:proteinase_dials}.

Across ten unit cells spanning all seven lattice systems, GLINT indexes
all four tested orientations for every cell, while its runtime remains
43--73~ms over a 19-fold range in reflection count
(Table~\ref{tab:cells}). DPS is complete on most of the tested cells
but indexes only 3/4 Proteinase K orientations, 2/4 of the large
orthorhombic case and 1/4 of the $140/140/150$~\AA\ tetragonal case.
Its runtime increases from 0.8 to 10.7~s across the same inputs,
corresponding to an $18$--$146\times$ timing difference. The largest
contrast occurs for the $140/140/150$~\AA\ cell, for which GLINT
indexes all four orientations in 73~ms while DPS indexes one of four in
10.7~s.

\section{Crystallographic validation and robustness}
\label{sec:validation}
\twocolumn

The preceding section established that GLINT recovers correct lattices at rates comparable with the strongest available blind and known-cell indexers, at substantially higher throughput. It remains to confirm that those assignments support downstream crystallographic analysis, and that the method stays selective outside the single-lattice setting. First, we carry GLINT solutions through
integration and merging \cite{crystfel,crystfel2} on several experimental serial data sets \cite{cxidb} and test
whether the recovered reflections retain crystallographic consistency.
Second, we test the recurrence criterion when multiple lattices are present
or when no crystal lattice is present at all.

\subsection{End-to-end validation on real data}
\label{sec:realdata}

End-to-end tests assess whether cells and orientations recovered by GLINT
support subsequent crystallographic processing. Merge quality is reported
using $CC^{*}$ \cite{ccstar}, $R_{\mathrm{split}}$ \cite{crystfel},
the mean reflection signal-to-noise ratio
$\langle I/\sigma\rangle$, and reciprocal-space completeness. \rev{In practice, merging statistics are used during acquisition at serial synchrotron beamlines to assess collection
progress \cite{basu2019}\rev{, and completeness alone is not a sufficiency criterion, since a nominally complete set is reached at lower multiplicity than well-determined intensities require \cite{vonstetten2026}}. All merges reported here, are evaluated offline after acquisition, and the streaming driver's running accumulator is not used in the analyses below.}

\paragraph{Jungfrau-4M lysozyme images.} GLINT performs peak finding, blind
cell determination, indexing and integration before external scaling and
merging with \texttt{partialator} \cite{partialator}. Of the 1563
frames, GLINT indexes 1506 (96\%) blind; 1482 (95\%) survive the
final acceptance gate and are merged. The resulting
1482-crystal data set reaches $CC^{*}=0.915$,
$R_{\mathrm{split}}=31.6\%$, $\langle I/\sigma\rangle=7.7$, and
99\% completeness at 2.1~\AA. A CrystFEL/XGANDALF processing of the
same frames, merged with the same \texttt{partialator} settings --- the integration step is not matched, since GLINT box-integrates its own predictions while the CrystFEL arm integrates in \texttt{indexamajig} --- gives $CC^{*}=0.930$ and
$R_{\mathrm{split}}=34.9\%$. Thus, the two summary statistics show
small differences in opposite directions: XGANDALF gives the higher
$CC^{*}$, whereas GLINT gives the lower $R_{\mathrm{split}}$. Both margins
are smaller than the measured sensitivity of these statistics to the one
unmatched setting: with solutions and \texttt{partialator} settings held
fixed, changing the CrystFEL integration radii (\texttt{-{}-int-radius})
from the $3,4,6$ used here to $1,2,4$ shifts $CC^{*}$ by $0.017$ and
$R_{\mathrm{split}}$ by $7.5$ percentage points on the same data set
(scored at $2.077$~\AA), so the head-to-head differences should not be
read as an accuracy ranking of the indexing engines.

\paragraph{cxidb-45 Proteinase K.}A different protein provides a stronger generalization test. On 907 readable cxidb-45 Proteinase K frames, GLINT recovers the tetragonal cell blindly and indexes 842 frames (93\%) from its uncapped peak lists. The corresponding integrated subset merges to $CC^{*}=0.90$ with $R_{\mathrm{split}}=31\%$ and $\langle I/\sigma\rangle=7.8$. In a controlled \textsc{nanoBragg} simulation, merges obtained from GLINT orientations and exact orientations are indistinguishable within the observed variation, suggesting that orientation error is not the dominant limitation of the merged intensity statistics in this test.

\paragraph{cxidb-17 lysozyme.} The matched GLINT processing yields
746 merged crystals, compared with 305 for XGANDALF, and increases
reciprocal-space completeness from 59.7\% to 79.5\%. The corresponding
half-dataset correlations are $CC_{1/2}=0.77$ and $0.73$, respectively.

\paragraph{Crystallographic validation of additional indexed frames.}
The higher indexing yield of GLINT raises a separate question: whether
the additional accepted frames contain coherent crystallographic signal
or arise from permissive lattice assignments. We therefore merged
GLINT-recovered subsets separately on several additional experimental
data sets. On cxidb-61 POMGnT1, the GLINT-recovered subset and an
XGANDALF-indexed control subset give $CC^{*}=0.848$ and $0.849$,
respectively. On cxidb-62, 2082 frames missed by XGANDALF but recovered
by GLINT merge to $CC^{*}=0.99$, and on cxidb-83 beta-lactamase,
590 of 600 such XGANDALF-missed frames merge to $CC^{*}=0.98$.
These independent merge controls are consistent with the additional
GLINT assignments containing coherent crystallographic signal rather
than arising solely from permissive lattice fits.

\paragraph{Merohedral
indexing ambiguities.} Additionally, cross-frame consensus determines the recurrent lattice but does not
resolve indexing-mode ambiguities that arise when the lattice symmetry
exceeds the symmetry of the diffraction intensities. Such merohedral
indexing ambiguities remain a downstream merging problem and are treated
by tools such as \texttt{ambigator} \cite{crystfel2,brehm2014} or
\texttt{dials.cosym} \cite{cosym}. Among the present validation sets,
this issue applies to cxidb-62; its effect is examined in Supporting
Information Sec.~S8.

\subsection{Multiple-lattice and non-crystal rejection}
\label{sec:multilattice}
The above consensus model assumes a recurrent lattice is present.
We further test two departures from that setting: \textit{overlapping crystals},
for which more than one orientation may contribute to a frame; and
\textit{non-crystal} input, for which no recurrent lattice should be returned.
The overlapping-crystal tests separately examine controlled recovery in
simulation and false second-lattice detections on experimental residual
peak lists.

For controlled two-crystal mixtures, we first test whether overlapping
peak sets can be separated without selecting a spurious lattice that fits
peaks from both crystals. Here QDIST scores a candidate using isotropic
peak-to-lattice distance in reciprocal space rather than the
fractional-index residual used by the standard single-lattice path.
After one lattice is accepted, \emph{deflate-and-reindex} removes peaks
assigned to that lattice and re-runs the indexer on the residual peak set.
The two operations are therefore complementary: QDIST selects the first
lattice more robustly in an overlapping pattern, whereas
deflate-and-reindex searches explicitly for an additional lattice.

\paragraph{Synthetic two-lattice mixtures.} With controlled relative intensities,
this procedure recovers both lattices in 96\% of cases when the second
lattice is at half intensity and 90\% when the two have equal intensity.
A stopping rule requiring a further lattice to explain at least 15 new
peaks and at least 20\% of the residual returns a median lattice count of
two without introducing a spurious third lattice in the tested mixtures.
Further detail is given in Supporting Information
Sec.~\ref{sec:si_multilattice}.

\paragraph{Real LCLS runs.} As a null control, each residual peak is independently rotated in azimuth about the beam axis, preserving its excitation error while destroying lattice coherence. On mfxl1038923 r0278 and r0058, without an additional physical gate, a second-lattice search succeeds on 95--96\% of the experimental residuals but also on 93--96\% of the
scrambled controls, showing that a residual containing many peaks almost
always admits an accidental lattice. Requiring the second solution to
share the first unit cell and to differ in orientation by at least
$15^{\circ}$ reduces the inferred double-hit fractions to 8.7\% and
8.9\% for the two runs, compared with 0\% for the corresponding
scrambled controls. Thus, in these data, the null comparison and
same-cell/orientation gates are required to interpret a second-lattice
solution as evidence of an overlapping crystal.

\paragraph{Non-crystal data.} The consensus vote also provides a direct non-crystal rejection criterion. Across 500 scramble-$q$ null groups, no leading cell satisfies the consensus acceptance gates; with
$N=3$ retained hypotheses per frame, the resulting crystal-versus-null
classifier has ROC AUC $0.993$. An independent crystal-free water
background run likewise produces no recurrent consensus cell. These
tests show that the same recurrence requirement used to recover weak
crystal lattices can also withhold a solution when the evidence is
incoherent across frames.

\section{Throughput engineering}
\label{sec:throughput}

Throughput engineering optimizes the complete execution path: before a unit cell is established, the
dominant cost is the blind M1--M6 per-frame solve (Fig.~\ref{fig:pipeline}); after consensus, the
steady-state workload becomes known-cell orientation registration. For online operation, peak finding and downstream frame processing add to this locked-state cost. We optimize these stages according to two principles: batch independent work across frames and keep intermediate data resident on the accelerator.

The sparse and dense front ends have different bottlenecks. \textit{Sparse stills} are dominated by the multi-start M3 refinement, whereas \textit{dense rotation} clouds spend most of their time in local-cluster extraction and FFT seeding. The M3 step-count sweep shows that the blind single-frame rate is flat within measurement noise above a few refinement steps, while the final consensus-plus-rescue rate is nearly unchanged over a much wider range. Reducing the shipped setting from 80 to 8 ascent steps therefore provides a substantial speed gain; re-measured at $n{=}480$, no step count from 4 to 80 detectably changes the final indexing yield. The analogous annealing sweep leads to the same conclusion: the cross-frame stage supplies most of the final robustness, whereas per-frame refinement primarily changes how many frames reach the faster path before rescue. Detailed stage profiles and parameter sweeps are given in Supporting Information Sec.~S4.

Once a consensus cell is available, known-cell registration becomes the
repeated steady-state indexing cost. Batching frames, capturing the GPU
stage as a CUDA graph, fusing the candidate operations and moving the
final cell check on device, and splitting the objective's candidates
across blocks, reduce this path from 2.4 to 0.17~ms per frame of throughput at 120 frames per batch on one
A100. The remaining scaling is primarily an occupancy effect: the anneal
and refine kernels take one thread block per frame, so throughput
improves with batch size. $B\simeq64$ is the streaming driver's default
but is not a saturation point --- $B=120$ measures $26\%$ faster --- and
the tabulated $0.17$~ms of throughput is measured at $B=120$, i.e. 120 frames per batch.
Kernel-level measurements are given in Supporting
Information Sec.~S4.

Cross-frame consensus also reduces the amortized discovery cost at the
scheduling level: once the cell has been established from an initial
subset of frames, the remaining patterns can enter the cheaper batched
known-cell path rather than being solved blindly. For example, on the 120-frame
cxidb-17, a consensus-first schedule requires 47 blind solves
instead of 120 while reaching a comparable final indexing yield ($94$ of
$120$ frames against $91$ for the blind-every-frame order; the
provenance of the three extra frames is given in Supporting Information
Sec.~\ref{sec:si_streamstress}). Thus,
for a long serial run, the blind discovery cost is set primarily by the
number of frames required to establish the cell rather than by the total
number of acquired frames.

\begin{table}
\caption{\label{tab:throughput}Throughput levers (one A100, sparse cxidb-17).}
\begin{tabular}{lrrr}
\toprule
lever & before & after & speedup \\
\midrule
M5 anneal: numpy loop $\to$ batched GPU solve & 2247 & 10 & $220\times$ \\
M3 ascent steps $80\to8$ (no detectable rate cost) & 95 & 23 & $4.1\times$ \\
M5 anneal iters $15\to3$ (over-annealed) & 7.8 & 1.6 & $4.9\times$ \\
known-cell rescue: numpy $\to$ GPU & 127 & 16.5 & $7.7\times$ \\
M2 de-dup $+$ M4 build: host $\to$ on-device & 11.6 & 3.2 & $3.6\times$ \\
N-best cell reduce: host loop $\to$ GPU dedup & 150 & 40 & $3.7\times$ \\
\midrule
\multicolumn{4}{l}{\emph{known-cell engine, batched across frames}} \\
coarse orientation grid (dirs/4) & 2.4 & 1.84 & $1.3\times$ \\
CUDA-graphed GPU stage & 1.84 & 1.46 & $1.26\times$ \\
fused CUDA kernels $+$ on-device cell check & 1.46 & 0.45 & $3.2\times$ \\
batch $32\to120$ (occupancy, same kernels) & 0.45 & 0.26 & $1.7\times$ \\
objective: candidates split across blocks (bit-exact) & 0.26 & \textbf{0.17} & $1.5\times$ \\
\midrule
\multicolumn{4}{l}{\emph{integration of the predicted reflections}} \\
whole-frame float64 upcast removed & 585 & 7.6 & $76.8\times$ \\
fused GPU box-integration (frame resident) & 7.6 & 0.33 & $23\times$ \\
\bottomrule
\end{tabular}

\footnotesize\noindent 
Times are in ms/frame. Rows in the upper block are sequential optimization
snapshots and should not be interpreted as current stage costs or summed
across rows. The known-cell block reports batched throughput on one A100
(fp64); the first three rows use batch size 32, and the final two show the
effect of increasing the batch size to 120 and splitting candidates across
blocks. Integration timings are for a 16\,Mpixel frame with 800 predicted
reflections. \rev{The graph and fused-kernel steps are bit-exact against the eager
path, the integration upcast removal is bit-identical, and the fused
integration kernel is bit-exact for integer and float32 detector data;
every step is rate-identical to within the discordant noise (S2).}
\end{table}

For the streaming driver, batched known-cell registration operates at 0.17~ms per frame of throughput at 120 frames per batch (about 5900 frames~s$^{-1}$) on one A100 GPU. Including peak finding and the surrounding locked-state pipeline gives 3.64~ms per frame (about 275 frames~s$^{-1}$), measured over a single pass of the 40-frame benchmark stack, lock-in frames included (the shipped driver default batch is 64 frames). These values are throughput operating points rather than single-frame latency. The reported speed ratios depend on whether a GPU is compared with a full CPU socket or with a single indexer instance; the normalization used for each head-to-head is stated with the corresponding benchmark.

\section{Outlook}
\label{sec:outlook}

\paragraph{Scope and current limitations.} The present implementation reports a reduced primitive cell, and the validation sets used here have primitive lattices; recovery of conventional centered settings is outside the scope of this evaluation. Cross-frame consensus determines a common lattice, but does not replace downstream treatment of indexing ambiguity or other crystallographic constraints. \rev{Consensus also has a finite resolving power in cell space: two lattices are pooled as one when their Buerger-reduced axis lengths agree to within $5\%$, the cosines of their reduced angles to within $0.06$, and their volumes to within $10\%$, so samples whose cells differ by less than this---polymorphs, or apo and ligand-bound forms of the same crystal---are indexed on the locked cell rather than reported as a change of sample. The complementary hazard is a derivative lattice of the locked cell rather than a merely similar one; in the streaming path it is suppressed by the pool-share and lead gates of Eq.~\eqref{eq:stream_lock}, the derivative-alias test being disabled at the shipped defaults.}

\paragraph{Extensions of the principle.} Blind indexing also improves when each exposure carries more reciprocal-space information: broader bandwidth, finite-crystal effects or moderate mosaicity thicken the region sampled near the Ewald sphere and reduce the single-shot degeneracy, until broadening exceeds the indexing tolerance (Supporting Information S6). Convergent-beam geometry extends each reflection into a Kossel streak \cite{cbxd}, and applying the same recurrence principle there is the subject of separate work. The position-only objective further suggests powder indexing, where orientation averages out and the problem becomes recovery of the reciprocal metric tensor from $d$-spacings, and Laue or pink-beam stills, which broaden the sampling at the cost of an unknown wavelength for each reflection (Supporting Information S9). Recurrence itself also generalizes: in cell space it identifies signal because the true lattice is common while aliases vary, whereas in detector space the sign reverses --- persistent artifacts recur at fixed positions while Bragg peaks move with orientation --- suggesting data-driven rejection of static artifacts (Supporting Information S14).

\paragraph{\rev{Consensus as a subject of study.}} \rev{The consensus mechanism itself invites further work. The phase diagram of Fig.~\ref{fig:stat_phase} reveals that past the generation limit, additional frames/pooling cannot restore hypotheses the per-frame search never produced. How the required pool size scales with spurious load, how quickly the sequential vote can be brought to lock, and whether a learned ranker can extend generation-side reach remain open questions.}

\paragraph{\rev{Streaming operation.}}
\rev{Further development could focus on tighter integration of GLINT with the
online analysis stack at high-repetition-rate facilities. In particular,
reducing the time required to establish a consensus cell would shorten the
initial blind-indexing stage before steady-state registration. Once a
cell is established, indexing is no longer the dominant computational cost,
suggesting that further throughput gains may come from integrating GLINT more
closely with upstream peak finding, including classical or learned approaches
\cite{peaknet}. Co-locating these operations within the facility data-reduction
layer, or ultimately closer to the detector \cite{thayer2025}, could provide a
path toward a fully integrated diffraction-to-indexed-data stream. Such system-
level integration is beyond the scope of the present work but is a natural
direction for future development.}

\section{\rev{Conclusions}}
\label{sec:conclusions}
\rev{Blind indexing of sparse serial stills is limited not by local refinement but by the accidental
lattice coincidences that a large candidate search produces on every frame --- a ceiling no
per-frame method can cross. GLINT crosses it by taking the data set rather than the frame as the
unit of inference, pooling weak but recurrent lattice hypotheses into a consensus cell against which
the remaining frames are registered; frames indexed this way merge to crystallographic quality
($CC^{*}=0.90$), and non-crystal images are refused rather than force-indexed. Blind cell
determination therefore moves out of offline processing and into the live data path, where the
result arrives while the experiment is still running.}

\section{Data and Code Availability}
\label{sec:repro}

GLINT is developed at \url{github.com/slac-lcls/glint}\rev{, and is packaged as a task for LUTE \cite{lute}, the workflow system used for automated analysis at LCLS}. The repository is currently private pending completion of SLAC's institutional software-release review; the source carries a BSD-3-Clause licence with a DOE enhancements grant-back, and it will be made public on completion of that review. Access for the editors and referees will be provided as soon as that review concludes. Unless noted otherwise, results were obtained with PyTorch 2.1.0 in the \texttt{ana-4.0.58-py3-minipytorch} environment on the S3DF A100 partition; the streaming per-stage timings were measured in \texttt{ana-4.0.59-py3-minipytorch} on the same hardware. Detailed script entry points and experiment-specific parameters are listed in the Supporting Information (S2). Public data-set identifiers and preprocessing details are given in the Supporting Information (S1). \rev{The remaining LCLS data sets (experiment cxil1015922, run 0033, and experiment mfxl1038923, runs 0278 and 0058) are not publicly deposited; they were used under LCLS's authorization for algorithm-development use of archived data, and requests for the underlying data should be directed to the facility and the experiments' PIs.}

\section{Acknowledgments}

\ack{
\vspace{5pt}
Work at SLAC National Accelerator Laboratory is supported by the U.S. Department of Energy
under Contract No.\ DE-AC02-76SF00515.
\rev{Use of the Linac Coherent Light Source (LCLS), SLAC National Accelerator Laboratory, is
supported by the U.S. Department of Energy, Office of Science, Office of Basic Energy Sciences
under Contract No.\ DE-AC02-76SF00515.}
\rev{Y.N.\ acknowledges support from the ILLUMINE project (Intelligent Learning for Light Source
and Neutron Source User Measurements Including Navigation and Experiment Steering), supported by
the U.S. Department of Energy, Office of Science, Advanced Scientific Computing Research program.}
Data handling and detector calibration build on the LCLS
\emph{psana} framework; GLINT's GPU calibration path is validated against psana's reference
implementation (M.~Dubrovin, SLAC/LCLS). \rev{We thank V.~Mariani and G.~Dorlhiac for
computational-environment, LUTE and DRP-pipeline setup that enabled the early stages of this
work; the teams of LCLS experiments mfxl1038923 (PI S.~G\"unther, DESY) and cxil1015922
(PI J.~Standfuss, PSI) for the archived data used, under facility authorization, to benchmark
the methods reported here; and V.~Mariani and A.~Tolstikova (DESY) for facilitating access to
the multi-hit runs. An independent GPU implementation of \texttt{peakfinder8} by G.~Dorlhiac
informed the iteration-count and precision choices in GLINT's vendored one.} Computations used
the SLAC Shared Scientific Data Facility (S3DF). This research used resources of the National
Energy Research Scientific Computing Center (NERSC), a U.S. Department of Energy Office of
Science User Facility, using NERSC award ERCAP0035513.

Claude (Anthropic) was used in assisting in developing the GLINT software, in the
analysis reported here and in preparing this manuscript. We used the large language model ChatGPT
by OpenAI to refine the language and enhance the readability of this
paper. The authors have verified
the results reported and take full responsibility for the content of this article.}

\clearpage
\setcounter{figure}{0}
\setcounter{table}{0}
\referencelist[glintNotes]

\global\def\thefigure{S\arabic{figure}}
\global\def\thetable{S\arabic{table}}

\clearpage
\onecolumn
\SItitle{Supporting information}
\makeatletter
\newcommand{\SIlabel}[2]{%
  \begingroup
  \def\@currentlabel{#1}%
  \label{#2}%
  \endgroup
}
\makeatother

\SIsec{S1. Data}
\noindent All public data sets used here are deposited at the Coherent X-ray Imaging Data Bank \cite{cxidb} (\url{https://www.cxidb.org}): entries 17 (doi:10.11577/1096920), 45 (doi:10.11577/1350027) and 62 (doi:10.11577/1365656) are distributed under the CC0 waiver; the beta-lactamase and POMGnT1 sets are CXIDB entries 83 and 61. The databank's stated aim --- community development of analysis tools for the anticipated torrents of XFEL data \cite{maia2016} --- is the use these benchmarks serve here.

\noindent\emph{cxidb-17 (CSPAD).} Peak lists are built at the deposition CrystFEL geometry with
\texttt{peakfinder8}; the 120-frame benchmark set is the first $120$ frames in deposition order
with no indexing-based selection. Reciprocal-space vectors for this set and its 480-frame
extension are computed at the fixed wavelength of the first frame, $\lambda=1.3222$~\AA, applied
to every frame and to all methods compared on these sets; the ${\sim}0.12\%$ per-shot wavelength
spread of the source is not applied. Peak lists for both sets were produced by CrystFEL 0.12.0
\texttt{indexamajig} with \texttt{--peaks=peakfinder8}, \texttt{--threshold=300},
\texttt{--min-snr=5} and \texttt{--min-pix-count=2}, all other peak-search parameters at their
CrystFEL 0.12.0 defaults; all methods compared on these sets consume the same lists.

\noindent\emph{cxidb-45 (SACLA MPCCD).} Eight-panel geometry, per-shot wavelength
($\lambda\!\approx\!0.95$~\AA); $907$ readable frames. \rev{Images are the deposited
\texttt{run296940-\{0,1\}.h5} stacks (the third deposited file does not read); the geometry is the
refined \texttt{mpccd-optimized.geom} of the cxidb-62 deposition (same MPCCD octal, $50$~$\mu$m
pixels, camera length $0.055$~m), and the wavelength is read per shot from each frame's
\texttt{photon\_wavelength\_A}. Peaks are $5\times5$ local maxima exceeding the image median by
$8\times$ the MAD-scaled robust $\sigma$, and the $170$ brightest per frame are kept
(\texttt{peakfind(img, nmax=170, snr=8.0)}); the same lists are given to both indexers in the DIALS
comparison.}

\noindent\emph{cxidb-62 / cxidb-83 / cxidb-61.} cxidb-62: ACG (\emph{Agrocybe cylindracea}
galectin), hexagonal $P6$, $105.8/105.8/75.5$~\AA, deposited hit set with its own stream peak
lists. cxidb-83: European XFEL
AGIPD, $600$ frames. cxidb-61: orthorhombic, ${\sim}800$ frames per arm.

\noindent\rev{\emph{Jungfrau-4M lysozyme (LCLS CXI, cxil1015922 run 0033).} Input is the run's four
Cheetah \texttt{.cxi} files ($404+373+409+377=1563$ frames). Peak lists are the Cheetah
\texttt{peakfinder8} lists stored in the files, reused unchanged; GLINT does not re-find peaks on
this set. Reciprocal-space vectors are computed through the refined CrystFEL geometry ($64$ panels,
camera length $0.111$~m) at the photon energy recorded in that geometry file ($8852$~eV,
$\lambda=1.4007$~\AA), applied to every frame. Blind indexing recovers the lysozyme cell on
$1506/1563$ frames and the acceptance gate retains $1482$, which are the frames merged in
Table~\ref{tab:realmerge}.}

\noindent\rev{\emph{mfxl1038923 runs 0278 and 0058 (LCLS MFX, Epix10ka2M).} Frames are read from the
LCLS xtc archive through psana; transverse pixel coordinates come from the psana geometry, while the
camera length is set per run ($0.060$~m for r0278, $0.050$~m for r0058) because the deployed psana
distance is a nominal placeholder, both anchored on the hexagonal-ice ring positions in the
frame-averaged radial profile. Wavelengths are taken per event from the electron-beam photon energy
(r0058) or fixed at $1.263$~\AA{} (r0278). Peaks are found with GLINT's local-annulus finder at its
Epix10ka2M defaults, and frames with fewer than six peaks are skipped. The per-frame hypothesis pools
of Sec.~S16 contain $1785$ (r0278) and $2319$ (r0058) indexed frames; the two runs return the same
lattice to within a few per cent per axis.}

\noindent\rev{\emph{Synthetic sets.} \emph{nanoBragg control (Sec.~\ref{sec:realdata}).} Forty
lysozyme stills ($P4_{3}2_{1}2$, $79/79/38$~\AA) were rendered with \textsc{nanoBragg}
(cctbx/simtbx) at uniformly random orientations on a single flat $1024^{2}$ panel ($0.2$~mm pixels,
$150$~mm, $\lambda=1.32$~\AA), with a $12\times12\times12$-cell Gaussian crystal shape and no
mosaic spread, divergence or spectral dispersion requested. Ground-truth amplitudes are random with a
$B=18$~\AA$^{2}$ falloff on the complete set to $d_{\min}=1.9$~\AA; images are scaled to $4000$
photons at the $99.99$th-percentile pixel of the first still, offset by a flat $8$-photon background
and Poisson-sampled with fixed seeds. Both arms share the prediction ($d_{\min}=1.9$~\AA,
excitation half-width $0.004$~\AA$^{-1}$) and box integration ($7\times7$ box, median background);
the exact arm takes the rendered orientation, the GLINT arm blind-indexes the $140$ brightest peaks.
Merging is CrystFEL~0.12.0 \texttt{process\_hkl} in $4/mmm$ with odd/even half-sets. Both arms
index $40/40$ and give $CC^{*}=0.860$ (exact) against $0.870$ (GLINT), $R_{\mathrm{split}}=57.2\%$
against $56.1\%$, and completeness $71.9\%$ in both. \emph{Dense-rotation reciprocal-lattice sets
(Table~\ref{tab:cells}).} For each cell, every reciprocal-lattice node with
$|\mathbf q|\le1/3$~\AA$^{-1}$ is generated exactly (no Ewald selection, positional noise or
intensities) at four uniformly random orientations with a fixed seed; the node count is the median
over the four. A solution counts when all three axis lengths are matched within $5\%$ (for DPS,
among its candidate lengths); times are medians over the four orientations.}

\SIsec{S2. Implementation Details}
\SIlabel{S2}{sec:si_impl}
\noindent\emph{Code paths.} Batch consensus is \texttt{glint.multishot}; the streaming driver is
\texttt{glint.stream\_driver}, with the buffered startup entering through \texttt{warmup\_batch};
the offline known-cell reference of Sec.~\ref{sec:streamresults} is \texttt{hybrid\_index} handed
the textbook cell with consensus skipped; the cascade fallback ships as \texttt{glint~-{}-cascade};
the oracle attribution of Sec.~\ref{sec:ceiling} is \texttt{oracle\_blind.py}; the spurious meter
and its per-run sweep ship as \texttt{glint.spurious\_meter} and \texttt{glint.spurious\_sweep}.

\noindent\emph{CrystFEL back-projection harness (head-to-head benchmarks).} Every indexer is run
inside one \texttt{indexamajig} harness on a synthetic flat panel whose recovered cells match the
input to $<1\%$ (CrystFEL 0.12.0, \texttt{-{}-peaks=cxi}, tolerances $5,5,5,1.5,1.5,1.5$; the cell
is supplied for the known-cell rows). XDS could not be added to this harness; the labelit/DPS
engine appears via its DIALS port. \rev{The standalone rows used DIALS 3.dev.1332 (git \texttt{3e144e7ea}; cctbx v2025.4-127) for the DIALS-FFT3D and DPS/rstbx rows, the fast-feedback indexer \texttt{ffbidx} at commit \texttt{a184a30} (26 July 2024) for the ffbidx row, and the \texttt{libxgandalf} shipped with CrystFEL 0.12.0 for the standalone XGANDALF row, the same shared object the in-CrystFEL path links, so both XGANDALF paths are one build. XGANDALF
was run at CrystFEL's default sampling settings (sampling pitch 6,
\texttt{denseWithSecondaryMillerIndices}; gradient-descent iterations 4, \texttt{manyMany}).}
\rev{The Proteinase~K comparison (Table~\ref{tab:proteinase_dials}) used the DIALS build of the \texttt{cctbx.xfel} container \texttt{dwpaley/cctbx-xfel:lz03-p2} on NERSC Perlmutter.}

\noindent\emph{Batched vs.\ per-frame known-cell registration.} The two paths agree to within the
discordant noise, but not per frame: they disagree at the acceptance gate on $18$ of $120$ and $52$
of $480$ frames ($9$ each way at $n{=}120$; $23$ vs $29$ at $n{=}480$, sign test
$p{=}0.49$)---the split is symmetric at $n{=}120$ and slightly favours the per-frame path at
$n{=}480$; neither is a systematic winner at this sample size.

\noindent\emph{Orientation refinement before merging (cxidb-17 full run).} Merging GLINT's raw
orientations pre-projection gives $CC_{1/2}{=}0.27$ unrefined; handing CrystFEL the raw orientation
and refining at loose tolerance in tPc before prediction is what the merge numbers use.

\noindent\emph{The $91$-vs-$92$ pair.} These are two pipelines at one gate, not one measurement
under two scoring conventions. $92/120$ is the blind pipeline, which derives its own cell by
cross-frame consensus; $91/120$ is the same registration path handed the reference cell with
consensus skipped (the known-cell GLINT-\textcircled{1} row of Table~\ref{tab:summary}). Each frame
is scored against the cell its own run used. The difference is not a fixed offset: it is one frame
at $n{=}120$ and four at $n{=}480$ ($361$ against $357$).

\noindent\emph{Null-width decomposition.} The pooled spurious-score width $\sigma{=}6.8$ mixes the
within-frame extreme-value fluctuation with frame-to-frame spread in peak count and spurious
load---at fixed frame composition the max-of-$K$ scale would be far narrower---so the histogram is
the operational null the meter tests against, not a pure max-of-$K$ law.

\noindent\emph{Prediction and integration.} GLINT predicts reflection positions from the indexed
orientation and box-integrates each one over a $7\times7$ pixel box against a surrounding annulus
($2$-pixel gap, $3$-pixel width), writes a CrystFEL-format stream, and hands it to
\texttt{partialator}. All intensities reported here use the annulus \emph{median} as the background
estimator, which is what GLINT shipped when these merges were made; the current default is a
MAD-clipped annulus mean, which removes a small positive bias in $I$ on sparse counting data. The
merges have not been re-measured under the new default, and the \texttt{median} setting reproduces
the values in Table~\ref{tab:realmerge}. \rev{Prediction keeps a reflection when its Ewald
excitation error $\zeta=q_z+\tfrac{\lambda}{2}|\mathbf q|^{2}$ satisfies $|\zeta|<\tau_E$; the
offline integrators and \texttt{glint -{}-int-tol} default to $\tau_E=0.006$~\AA$^{-1}$, while
the streaming driver applies the same test with $\tau_E=0.002$~\AA$^{-1}$, so its per-frame
reflection lists are narrower than the offline ones.}

\noindent\rev{\emph{Streaming-driver geometry and symmetry.} The streaming driver takes one
wavelength and one detector distance at construction and applies them to every frame: the
peak-to-$\mathbf q$ mapping, the reflection grid and the prediction all use these fixed values,
and \texttt{push()} receives only the image. The offline \texttt{-{}-images} route instead reads
\texttt{clen} and \texttt{photon\_energy} per event from the CXI file when the geometry gives
them as HDF5 paths. Live geometry refinement, when enabled, is diagnostic only and is not fed back
into prediction. The running merge statistics the driver reports (completeness, $CC_{1/2}$,
$CC^{*}$, $R_{\mathrm{split}}$) are accumulated in Laue group $4/mmm$ after the locked cell is
standardized to a tetragonal setting with the unique axis along $c$; they are therefore meaningful
only for tetragonal cells such as lysozyme, and every merge statistic reported in this paper is
evaluated offline (Sec.~\ref{sec:realdata}) without the live accumulator.}

\noindent\emph{Peak-list caps.} No peak-list cap is applied by default: the shipped settings are
uncapped (\texttt{M3\_CAP}${=}0$, \texttt{-{}-top-peaks}${=}0$), and every GLINT rate reported here
is measured on uncapped peak lists. An optional setting feeds at most $N$ peaks, taken in the
frame's own order, to the M3 ascent alone while M2 and M4--M6 retain the full list; at $N{=}100$
this runs ${\sim}1.3\times$ faster at an essentially unchanged rate ($84/80$ of $120$ against the
$85/79$ baseline of Sec.~\ref{sec:ceiling}), and selecting by intensity instead measured worse
($82/78$). Truncating the peak list at ingest is a different operation and is measurably harmful
($71/66$), so it is not used. The only capped lists in this work are the $170$-peak Proteinase~K
lists of the same-input DIALS comparison (Sec.~\ref{sec:realdata}).

\SIsec{S2a. Module architecture (M1--M6)}
\SIlabel{S2a}{sec:si_modules}

\paragraph{M1: candidate-vector initialization.}

In the absence of a supplied unit cell, GLINT must initialize both the
direction and the length of candidate real-space lattice vectors. Let
$\left\{
\widehat{\mathbf u}_{j}
\right\}_{j=1}^{N_{\mathrm{dir}}}
\subset \mathbb{S}^{2}$
denote a near-uniform set of directions generated by Fibonacci-sphere
sampling (FSS) \cite{gonzalez2010fib,hannay2004fib}, and let
$\left\{
\ell_{s}
\right\}_{s=1}^{N_{\mathrm{len}}}$
denote the sampled vector lengths. 

M1 forms the initial candidate set
\[
\mathcal{V}_{0}
=
\left\{
\ell_{s}\widehat{\mathbf u}_{j}
:
1\leq j\leq N_{\mathrm{dir}},
\;
1\leq s\leq N_{\mathrm{len}}
\right\}.
\]
For the sparse-still configuration used here, the length shells span
$30$--$126$\,\AA.

This construction replaces the ``lengths given'' initialization of the
known-cell fast-feedback procedure \cite{ffbidx}. Its purpose is not to
identify a cell directly, but to place starting points throughout the
physically relevant real-space search volume from which candidate lattice
vectors can be located by the periodic objective in M2--M3.

\paragraph{M2: direct-sum lattice-vector score.}

Each candidate vector $\mathbf v\in\mathcal{V}_{0}$ is evaluated using
the grid-free direct-sum objective
\begin{equation}
\begin{split}
\label{eq:direct_sum}
    \Phi_f(\mathbf v)
=
\sum_{i=1}^{N_f}
w_{fi}\,
\mathbf 1\!\left[
\left|
\mathbf q_{fi}\cdot\mathbf v
-
\operatorname{round}\!\left(\mathbf q_{fi}\cdot\mathbf v\right)
\right|
<
\epsilon
\right]
\cos\!\left(
2\pi\,\mathbf q_{fi}\cdot\mathbf v
\right),
\end{split}
\end{equation}
with $w_{fi}
=
\frac{1}{\|\mathbf q_{fi}\|},$ following the real-space scoring construction used by XGANDALF
\cite{Gevorkov2019XGANDALF}. If $\mathbf v$ is a real-space lattice
vector, the Laue condition of Sec.~\ref{sec:indexing_problem} places
$\mathbf q_{fi}\cdot\mathbf v$ near an integer for lattice-consistent
peaks, so their contributions add coherently. Incompatible vectors
produce phases distributed away from integer values and therefore a
smaller coherent sum. Normalization by $1/\|\mathbf q_{fi}\|$ reduces the disproportionate influence
of high-resolution reflections. The indicator $\mathbf 1[\cdot]$ is a
hard proximity window: only peaks whose value of
$\mathbf q_{fi}\cdot\mathbf v$ lies within $\epsilon$ of an integer
contribute, so peaks inconsistent with $\mathbf v$ neither add to the
score nor pull its gradient. At the shipped settings $\epsilon=0.18$,
the XGANDALF inlier tolerance. 

GLINT additionally uses a
$\cos^{2}$-sharpened form during candidate discrimination to narrow the
periodic maxima; at the shipped settings the sharpened form is applied
throughout. Because the phases are evaluated directly at the
candidate vectors, no reciprocal-space gridding is required. The
relationship of this objective to Fourier, Patterson and projection-based
candidate generation is developed in Sec.~\ref{sec:lineage}.

\paragraph{M3: continuous lattice-vector refinement.}

The discrete M1 samples are only initial guesses. For each seed
$\mathbf v^{(0)}$ in the M1 bank, M3 solves the local, unconstrained
optimization problem
\[
\widehat{\mathbf v}
=
\underset{\mathbf v\in\mathbb{R}^{3}}{\operatorname{arg\,max}}
\;
\Phi_f(\mathbf v),
\]
initialized at $\mathbf v^{(0)}$. Note the ascent direction is the
gradient of the direct-sum objective, with the indicator of
Eq.~\eqref{eq:direct_sum} treated as locally constant:
\[
\nabla \Phi_f(\mathbf v)
=
-2\pi
\sum_{i=1}^{N_f}
w_{fi}\,
\mathbf 1\!\left[
\left|
\mathbf q_{fi}\cdot\mathbf v
-
\operatorname{round}\!\left(\mathbf q_{fi}\cdot\mathbf v\right)
\right|
<
\epsilon
\right]
\mathbf q_{fi}
\sin\!\left(
2\pi\,\mathbf q_{fi}\cdot\mathbf v
\right).
\]
The window is re-evaluated at every step, so a peak enters or leaves
the active set as $\mathbf v$ moves.
GLINT uses the cosine-gradient refinement of the fast-feedback
cell-assembly procedure \cite{ffbidx}, using momentum to move each
initial seed toward a nearby continuous maximum \rev{(using eight ascent steps in the shipped default)}.

The role of M3 is therefore local rather than combinatorial: M1 provides
coverage of the search space, while M3 removes the discretization of that
initial sampling and converts neighbouring starts into estimates of
continuous real-space lattice vectors. The effect of the iteration count
on the single-frame rate and on computational cost is treated in
Sec.~\ref{sec:throughput}.

\paragraph{M4: candidate-basis assembly.}

The output of M3 is a set of refined candidate lattice vectors 
\[
\widehat{\mathcal V}_f
=
\left\{
\widehat{\mathbf v}_{1},
\ldots,
\widehat{\mathbf v}_{K_f}
\right\},
\]
where $K_f=|\widehat{\mathcal V}_f|$ is the number of refined
candidate lattice vectors retained after M3. Retention de-duplicates the
converged maxima on a $2$\,\AA{} real-space grid, discards vectors
shorter than $20$\,\AA, and keeps at most the $30$ highest-scoring
survivors, so $K_f\leq30$ and the enumeration below is bounded by
$\binom{30}{3}=4060$ triplets however many maxima M3 produces; the
sensitivity of the blind rate to this cap is reported in Supporting
Information Table~\ref{tab:negatives}. M4 subsequently forms candidate real-space bases by selecting linearly independent triplets (all $\binom{K_f}{3}$ combinations are formed, and near-coplanar triplets are removed by the relative-volume gate below, so independence is enforced per triplet rather than by a separate maximal-independent-set selection; a frame with $K_f<3$, or with no triplet passing the gate, contributes no basis hypothesis),
\[
M_{jkl}
=
\begin{pmatrix}
\widehat{\mathbf v}_j &
\widehat{\mathbf v}_k &
\widehat{\mathbf v}_l
\end{pmatrix},
\qquad
j<k<l,
\]
with $|\det M_{jkl}|
\geq
0.1\,
\|\widehat{\mathbf v}_j\|
\|\widehat{\mathbf v}_k\|
\|\widehat{\mathbf v}_l\|.$
By construction, each $M_{jkl}$ is therefore a complete candidate for the oriented
real-space basis defined in Sec.~\ref{sec:indexing_problem}.

This step is specific to the blind path. In fast-feedback indexing the
unit-cell dimensions are supplied, so candidate orientations can be
constructed by rotating a known basis \cite{ffbidx}. In GLINT the basis
itself is unknown, and M4 instead assembles independently recovered
real-space vectors into candidate cells before joint refinement.

\paragraph{M5: residual-threshold cell refinement.}

A basis assembled from three independently refined vectors is generally
not yet jointly optimal for the observed peak set. M5 therefore refines
each candidate basis by alternating Miller-index assignment and
least-squares refitting. For a current basis $M^{(t)}$, each observed
peak is assigned the nearest integer triplet
\[
\mathbf h_{fi}^{(t)}
=
\operatorname{round}
\!\left(
\bigl(M^{(t)}\bigr)^{T}\mathbf q_{fi}
\right),
\]
with residual
\[
r_{fi}^{(t)}
=
\left\|
\bigl(M^{(t)}\bigr)^{T}\mathbf q_{fi}
-
\mathbf h_{fi}^{(t)}
\right\|_{\infty}.
\]
At iteration $t$, only reflections satisfying the current residual
threshold $\tau_t$ are retained,
\[
\mathcal I_t
=
\left\{
i:
r_{fi}^{(t)}\leq\tau_t
\right\},
\]
and the basis is updated by solving
\[
M^{(t+1)}
=
\underset{M}{\operatorname{arg\,min}}
\sum_{i\in\mathcal I_t}
\left\|
M^{T}\mathbf q_{fi}
-
\mathbf h_{fi}^{(t)}
\right\|^{2}.
\]
The threshold is contracted geometrically,
$\tau_{t+1}=\max(0.85\,\tau_t,\,0.02)$ from $\tau_0=0.25$, so that the
fit is initially tolerant of an imperfect assembled basis and becomes
progressively restricted to lattice-consistent reflections; at the three
iterations used here the schedule runs $0.25\to0.2125\to0.181$ and does
not reach the floor.

This rounding-and-refitting operation is the iterated refinement used in
the fast-feedback indexing procedure \cite{ffbidx} and subsequently in
TORO \cite{toro}. GLINT applies the same crystallographic update to the
blindly assembled candidate cells, but evaluates the independent
candidate fits as a batched GPU operation using masked normal equations.
The current implementation uses three annealing iterations; the batching
and iteration-count choices are discussed in Sec.~\ref{sec:throughput}.

\paragraph{M6: candidate selection and lattice reduction.}

The refined candidates from M5 must finally be ranked. A score based only
on residual tightness is insufficient for sparse blind indexing, because
a spurious lattice can fit a small subset of peaks extremely well while
explaining little of the observed pattern (Supporting Information Table~\ref{tab:negatives}). GLINT therefore couples
residual quality to peak coverage.

For a candidate basis $M$, define its indexed set at the selection
tolerance $\tau=0.15$ --- a fixed value, independent of the M5 schedule,
and the same $\tau$ used in Eq.~\eqref{eq:stream_frame_accept} --- as
\[
\mathcal I_f(M)
=
\left\{
i:
\left\|
M^{T}\mathbf q_{fi}
-
\operatorname{round}(M^{T}\mathbf q_{fi})
\right\|_{\infty}
\leq\tau
\right\},
\]
and its peak coverage as
\[
c_f(M)
=
\frac{|\mathcal I_f(M)|}{N_f}.
\]
Among candidates satisfying the required coverage
($c_f(M)\geq0.30$ at $\tau=0.15$, with a floor of at least eight
inliers), M6 ranks the cells by
their lattice defect,
\[
D_f(M)
= \frac{1}{3|\mathcal I_f(M)|}
\sum_{i\in\mathcal I_f(M)}
\sum_{c=1}^{3}
\left|
(M^{T}\mathbf q_{fi})_c
-
\operatorname{round}\!\left((M^{T}\mathbf q_{fi})_c\right)
\right|.
\]

The coverage term dominates the ranking, so any candidate clearing the gate outranks every candidate that does not, and no candidate can win on a handful of exceptionally tight residuals alone. If no candidate clears the gate, the tightest remaining candidate is returned and is filtered by the cross-frame consensus rather than accepted on one frame's evidence. This coverage-gated selection is the final blind-cell discriminator before
cross-frame consensus. Empirical evidence of the coverage gate selection is given in Supporting Information Table~\ref{tab:negatives}.

The selected basis is subsequently reduced to a primitive representation
before cell comparison and consensus. The reduction selects the three shortest independent vectors found by a
bounded integer search over the assembled basis (coefficients in $[-3,3]$),
returning the basis unchanged when no such triple exists; a data-driven step
additionally halves any real axis whose inlier Miller indices are uniformly
even, reverting if that halving loses inliers. This is a shortest-vector
rather than a Niggli reduction, but the sorted reduced parameters are
invariant enough in practice that lattice equality can be tested by comparing
them directly --- the grouping tolerances of Sec.~\ref{sec:algorithm} operate on
exactly these reduced parameters --- and it is forced by M4's construction,
since bases assembled from unordered candidate axes arrive in no particular
setting. Consequently, a centred conventional
cell is represented internally by its primitive lattice, and cell
comparisons are performed between reduced representations. Recovery of a
particular conventional centred setting is not part of the present
implementation.

Together, M1--M6 constitute the per-frame engine of Fig.~\ref{fig:twopaths}: the blind path runs
the full sequence to emit the $N$-best hypotheses that the cross-frame consensus of
Sec.~\ref{sec:algorithm} pools, while the known-cell path shares the M3 cosine-ascent refinement
and the M5 annealing update but replaces M4 and M6 with a rotation search at the supplied cell and
an inlier/trimmed-log2 ranking.

\movedoff

\SIsec{S3. Candidate-generation lineage}

The winner and the reason it won are stated in the main text (Sec.~\ref{sec:lineage});
Table~\ref{tab:lineage} catalogues every representation-level candidate generator we built and
tested on the way there.

\makeatletter
{\singl@coltrue
\begin{table}
\caption{\label{tab:lineage}Representation-level candidate generators we tried. All are
the same signal (Wiener--Khinchin); the gridless direct sum with continuous ascent won.}
\begin{tabular}{p{2.5cm}p{6.6cm}p{6.7cm}}

generator & idea & verdict \\
\midrule
gridded 3D FFT (cuFFT) & deposit cloud on a grid, FFT, peak-find $|F(x)|$ & aliasing $+$ CIC digitization error $\propto\!\delta q\,x$ kills the long axis \\
difference / Patterson cloud & long axis $a\!\to\!$ short $a^{*}{=}1/a$; $P^2$ diffs pile on recip.\ lattice & single-shot degenerate: Ewald slice rank-deficient ($\sigma\!\approx\![1,0.88,0.16]$) \\
cropped autocorrelation & band-limit $|F|^2$ $\Rightarrow$ coarse alias-free grid, large extent & works, but no better than the direct sum \\
NUFFT (cuFINUFFT, KB kernel) & $|F(x)|$ at FFT speed $+$ direct-sum accuracy; peak-count-independent & fixes digitization but does not beat direct-sum ascent \\
DPS / projection-slice (MOSFLM) & 1D periodicity per direction; robust on sparse frames & complements only: ceiling $+2\%$, solve $-7\%$ \\
\textbf{direct sum $+$ continuous ascent} & exact phases, no grid; ascend to continuous maxima & \textbf{winner $\to$ GLINT M1--M3} \\

\end{tabular}
\end{table}
}

\emph{The partial-rotation continuum.} Between the single still and the full sphere lies the
oscillation wedge, the common rotation-data case. Across six cells spanning the lattice systems
(lysozyme, Proteinase~K, thaumatin, hexagonal, cubic, and a large orthorhombic cell; three random
orientations each) swept from a still slice through $20^{\circ}$ and $60^{\circ}$ wedges, the local-cluster
front end indexes $100\%$ at every wedge, \emph{flat at $\sim$60~ms} regardless of cell or reflection
count ($10^{3}$--$4\times10^{4}$). Blind Fibonacci matches this only on the small, compact lysozyme cell
(where both hold $100\%$ to a $60^{\circ}$ wedge and the gain is pure \emph{speed}---the blind objective
cost grows with the reflection count, $60\to820$~ms as the wedge widens, versus the flat $\sim$56~ms
cluster, a $15\times$ speed-up at matched accuracy). On the larger or more anisotropic cells the blind
search degrades in \emph{both}: its cost balloons past a second and its rate collapses by a $20$--$60^{\circ}$
wedge (thaumatin and orthorhombic already fail at $20^{\circ}$; cubic and hexagonal by $60^{\circ}$), while
the cluster front end stays at $100\%$/$\sim$60~ms throughout. The single still slice is degenerate for both
(the thin-Ewald SFX regime, recovered by cross-frame consensus, main text). Thus, through the
practical oscillation-wedge range, the local-cluster front end is flat and complete across cells. Blind
indexing loses speed on every cell and accuracy on all but the smallest.

\emph{The shared signal is a Dirac comb.} Underneath, that shared signal is a \emph{Dirac comb}: a crystal's lattice Fourier-transforms to
the reciprocal comb, which is why every reflection sits at $h\,a^{*}+k\,b^{*}+l\,c^{*}$, labelled
by three integers. Every generator above performs one
operation, the recovery of that comb's \emph{period}, and that recovery is robust on two counts: reading only positions,
it is invariant to the binarization that thresholded peak-finding imposes; and a sparse, partial
comb still fixes the period the way a few teeth fix a whole comb's spacing. That tolerance of
undersampling has a floor, since too few real peaks among spurious ones defeat single-frame
recovery. However, successive stills sample the same reciprocal comb, so the period that recurs
across them is the true cell (Sec.~\ref{sec:consensus}).

\emph{Why unit weights.} Every method in this lineage reads peak \emph{positions} and discards
\emph{intensities}. Each reflection enters with unit weight, and indexing works because of that
binarization.
The amplitude factors as $F(q)=F_{\mathrm{cell}}(q)\sum_n e^{2\pi i\,q\cdot R_n}$. The lattice sum fixes
\emph{where} the peaks sit (the geometry we recover), and the molecular transform $F_{\mathrm{cell}}$
fixes only \emph{how strong} they are. So the unit-weight comb $\sum_i\delta(q-q_i)$ transforms to the sharp direct lattice,
whereas the intensity-weighted comb $\sum_i I_i\,\delta(q-q_i)$ transforms to that lattice convolved with
the cell's Patterson, smearing the very peaks being located; unit weights deconvolve the contents away and
leave pure periodicity. Any weight restored afterwards (our $1/|q|$, xgandalf's $1/|q|^2$, a resolution
taper) is a function of $|q|$ alone, so it only rebalances resolution shells; the structure
factor stays deconvolved away. GLINT's objective $\sum_i w_i\cos(2\pi\,x\cdot q_i)$ inherits this: $w_i$ is geometric and the
peak list intensity-agnostic.

\emph{Why not a learned dictionary.} Every generator above is a \emph{fixed} transform (the prechosen-basis pole), as the DCT, wavelets
and curvelets are in image processing. The opposite pole learns an overcomplete dictionary from the
data itself (K-SVD \cite{ksvd}), and one could in principle replace the transform with such a learned
representation. We do not, because the indexing prior is a \emph{group} and not generic sparsity over a
dictionary. The reflections are the integer lattice fixed by three basis vectors (nine real
numbers), so there are no atoms to estimate, and a learned dictionary would spend its capacity
re-discovering the lattice structure the transform already encodes exactly. Learning belongs
upstream of the three-vector fit, in denoising the raw image or in triaging powder profiles for phase
identification.

\emph{Consensus as a stability-selection rule.} The ensemble template of the main text (Sec.~\ref{sec:consensus}) is a statistical selection rule, and GLINT's cross-frame consensus step---choosing the cell that the most frames recur on---applies it. (Within a frame, converged ascent maxima are de-duplicated by position and ranked by objective value rather than by how many starts reach them, so the single-frame stage is a selection, not a second recurrence vote.) Selecting the cell that
minimizes one frame's residual is the indexing analog of choosing a model by cross-validated fit, and
it over-selects for the same reason. On sparse, spurious-laden peak lists a false lattice can fit one
frame's positions by coincidence. Selecting by agreement across resampled data instead corresponds to the estimation-stability and
stability-selection criteria of high-dimensional regression \cite{escv,stabsel}, which cut false
positives sharply at little cost in true positives. The same consensus that raises the rate therefore also
refuses non-crystals: sensitivity and specificity are read from one stability statistic
(Sec.~\ref{sec:multilattice}). Serial crystallography supplies the resampling for nothing, because
successive stills of a running crystal are independent draws of the same reciprocal comb. Thus the stable
cell emerges with no bootstrap or held-out split.

\SIsec{S4. Throughput: per-stage detail}

\begin{figure}
\caption{The M3 step count bounds the blind front-end time, while the final rate stays flat. Per-stage time (stacked:
M1--M3 ascent vs the M4--M6\,$+$\,host tail) and two rates---\emph{blind} (single frame) and after
cross-frame \emph{consensus}$+$rescue---versus M3 ascent steps, one A100 over 120 cxidb frames. The
ascent grows linearly while the anneal-and-select tail stays flat ($\sim$6\,ms). Both rates here
are scored at the correct-lattice criterion, not the strict $\geq$25\%-of-spots bar of
Table~\ref{tab:summary}. The blind rate
is flat within measurement noise from a few steps upward (dashed, the shipped point) and collapses to 30\% at zero steps, yet the
consensus rate holds $\sim$96--98\% throughout. End-to-end $\sim$26\,ms/frame amortized
(Table~\ref{tab:stages}). Re-measured at $n{=}480$, no arm from 4 to 80 differs detectably from 8; any undetected gain is bounded jointly at $+7.4\%$ blind (95\%).}
\includegraphics[width=\columnwidth]{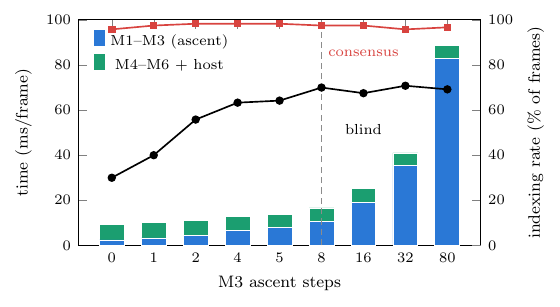}
\label{fig:steps}
\end{figure}

The two regimes have opposite time profiles (Table~\ref{tab:stages}), and the blind
front-end cost is set almost entirely by the M3 ascent step count (Fig.~\ref{fig:steps}).

\twocolumn
\begin{table}
\caption{\label{tab:stages}Per-stage time, sparse vs dense (one A100, shipped defaults
$\mathrm{STEPS}{=}8$, anneal${=}3$): the blind still path (120 cxidb frames) vs a dense
rotation cloud (52\,ms/cloud, ten cells). The bottleneck inverts (bold). Once the anneal (M5) is cut to
three iterations, sparse is dominated by the $70{,}400$-start ascent (M3, 58\%) and dense by the
cluster-FFT seeding (M1, 71\%). The upper block is the blind single-frame pipeline; the lower block adds
the two cross-frame stages that carry the correct-lattice rate to $\sim$97\%; at the strict
$\geq$25\%-of-spots bar of Table~\ref{tab:summary} the same pipeline gives $92/120$
(Fig.~\ref{fig:steps}).
$^{a}$known-cell index (rescue), median 17\,ms, run on the 35/120 frames (29\%) that fail blind
$\Rightarrow$ $17{\times}35/120\approx5$\,ms/frame amortized. $^{b}$multi-hypothesis cross-frame
consensus, once per run ($0.16$\,s${/}120\approx1.3$\,ms/frame). Dense clouds self-index, so neither
applies; end-to-end sparse $\approx26$\,ms/frame (the shipped $N$-best pipeline). $^{c}$optional self-contained front end: GLINT's
vendored GPU peakfinder (\texttt{peakfinder\_v4}/\texttt{9}) on real CSPAD frames, same A100 (the
CrystFEL/Cheetah-standard \texttt{peakfinder8} \cite{crystfel,cheetah} is $\sim$4.3\,ms). The benchmark reuses each frame's stored
\texttt{peakfinder8} peaks, so M0 is outside the peaks-given totals; adding it gives raw image
$\to$ indexed $\approx$29\,ms/frame. $^{d}$measured end-to-end wall-clock of the shipped $N$-best pipeline (main text); it exceeds the single-best per-stage sum plus rescue and consensus because the default retains the top-3 hypotheses.}
\begin{tabular}{lrrrr}
\toprule
 & \multicolumn{2}{c}{sparse (still)} & \multicolumn{2}{c}{dense (rotation)} \\
\cmidrule(lr){2-3}\cmidrule(lr){4-5}
stage & ms & \% & ms & \% \\
\midrule
M1 seeds    & 0.1 & $<$1        & 36.6 & \textbf{71} \\
M2 score    & 2.1 & 14          & 8.9  & 17 \\
M3 refine   & 8.6 & \textbf{58} & 2.4  & 5 \\
M4 assemble & 1.0 & 7           & 1.1  & 2 \\
M5 anneal   & 1.6 & 11          & 1.7  & 3 \\
M6 select   & 1.5 & 10          & 1.2  & 2 \\
\midrule
blind / frame          & 15          & 100 & 52  & 100 \\
\;$+$ rescue$^{a}$      & 5.0         &     & --- &     \\
\;$+$ consensus$^{b}$   & 1.3         &     & --- &     \\
end-to-end / frame$^{d}$ & \textbf{26} &     & 52  &     \\
\midrule
\;$+$ peak-find (M0)$^{c}$ & 3.2      &     & --- &     \\
raw image $\to$ indexed & \textbf{29} &     & --- &     \\
\bottomrule
\end{tabular}
\end{table}

\onecolumn

\emph{The batched known-cell engine, kernel by kernel.} Once consensus has fixed the cell, the
steady-state cost is the known-cell engine, and there the levers are different in kind. Because the
frames are independent, they can be batched. Batching removes the per-frame Python and lets one pass
cover many frames. What remains is then dominated first by host kernel dispatch (thousands of small
launches per batch) and, once those are collapsed, by the launch and execution overhead of the many
small tensor operations themselves. Capturing the whole GPU stage as a CUDA graph removes the
dispatch cost (a single replay per batch), which an analytic $3\times3$ solve/determinant makes
possible at all, since cuSOLVER forces a stream synchronisation and cannot be captured. Rewriting
the per-candidate hot loops (the objective, the gradient refine and the cell anneal) as fused CUDA
kernels then removes the small-operation overhead: one thread block per frame, one thread per
candidate, that frame's peaks staged once in shared memory, and the $3\times3$ normal equations and
their closed-form solve kept in registers. Moving the last host-side step (a per-frame reduced-cell
comparison) on-device as a batched reduction leaves only the final cell per frame to copy back.
Together these take the known-cell engine from $2.4$ to $0.17$\,ms/frame of throughput at 120 frames per batch (Table~\ref{tab:throughput},
main text) with the indexing rate unchanged in aggregate. At fp64 the batched and unbatched paths
agree to within the discordant noise: they gate differently on $18$ of $120$ and $52$ of $480$
frames, split $9$/$9$ and $23$/$29$ (sign test $p{=}0.49$), so neither is a systematic winner, but
the totals are not identical at $n{=}480$ (Table~\ref{tab:summary}). The
remaining limit is occupancy, and it is only partly removable. One block per frame means the batch
size sets how many multiprocessors are busy. The objective is the exception: it scores several
thousand candidates per frame against a $128$-thread block, so each thread walked $32$--$45$ of them
in series. Giving the candidate axis its own grid dimension turns a $120$-frame batch from $120$
blocks into $3840$--$5400$ and is bit-exact --- it re-maps which thread owns which candidate and
leaves every summation order untouched, so the per-frame output is identical --- for $1.5\times$ at
$B{=}120$ (fp64; the fp32 path, already less objective-bound, gains $1.1\times$).

That does not remove the batch. The anneal and the gradient refine still take one block per frame,
and measured across $B$ the engine still improves through to the full batch ($0.33$\,ms of throughput at 32 frames, $0.17$ at 120, the one exception being the ragged-batch tie noted below), with the spread
across $B$ widening rather than narrowing once the objective is split. We therefore quote $B{=}120$, the batch at which the tabulated figure was measured, and
note that the apparent plateau near $B{\simeq}64$ is an artefact of the benchmark set: $120$ frames
at $B{=}96$ is a ragged $96{+}24$ while $B{=}120$ is a single batch, so that part of the curve
reflects batch-count quantisation rather than saturation.

\emph{Integration: a profile-first story.} Once indexing costs a fraction of a millisecond the next
stage sets the pace, and the same profile-first discipline applies to it. Integration's cost lay entirely outside the
kernel. The routine opened by upcasting the whole detector frame to double precision
on every call, which is $O(HW)$ work for $O(n_{\mathrm{refl}})$ of integration. On a 16\,Mpix frame
that cast alone was $\sim$546 of the 585\,ms, and the whole-frame allocations it forces carried most
of the rest. The total barely moved when the reflection count went
from 800 to 2000, so the cost was not proportional to the work being asked for. Gathering each box
in the detector's native dtype and promoting only the boxes is bit-identical and takes the same
frame to 7.6\,ms. A fused GPU box-integration then reaches $0.33$\,ms per frame of integration, flat in both frame
size and reflection count. That last figure holds under one condition. A host-to-device copy of a
16\,Mpix frame costs $\sim$3--5\,ms, an order of magnitude more than the kernel, so the fused kernel
pays only when the frame is already resident on the device. That is the online case, where detector
data reach GPU memory directly and the peak search has already run there.

\emph{One engineered exception to the streaming driver's completeness.} A synthetic
severe-spurious-load stress test surfaced one exception to the claim that what the streaming driver
leaves unindexed is genuinely unindexable (main text, Sec.~\ref{sec:streaming}): the registration
engine's axis-candidate deduplication could itself discard the correct orientation under enough
contamination, before the anneal ever saw it. Tightening the deduplication radius closes
it---$494\to600/600$ known-cell solves at the worst synthetic point, a negligible $2/600$ regression
at the mildest load tested and none elsewhere---with no change to the known-cell rate in
Table~\ref{tab:summary}, whose offline evaluation already tolerated the gap via a per-frame blind
fallback the live fast path does not have.

\SIsec{S5. Single-frame levers explored}

Blind indexing on sparse cxidb saturates at $\sim$71\% of frames carrying a correct lattice
($85/120$; $66\%$ at the $\geq$25\%-of-spots bar) because the data is
\emph{spurious-limited}: richer candidates, peaks, or scorers add spurious signal faster than
true signal. Every single-frame lever we tried fails for that one reason
(Table~\ref{tab:negatives}).

\SIsec{S6. Reverse-space scoring and the fat Ewald slice}

\noindent\emph{Non-monotonicity of the single-frame rate in mosaicity (Fig.~\ref{fig:stat_phase}a).}
At $\sigma{=}0$ the curves start on the pure spurious floor ($100\!\to\!57\%$ as
$f\!:\!0\!\to\!0.8$), and $\mathrm{rate}_1$ is non-monotonic in $\sigma$: a small broadening
raises it, peaking near $\sigma\!=\!0.125\times10^{-3}$~\AA$^{-1}$, and the lift is largest where
the floor is lowest---at $f{=}0.8$, $57\!\to\!97\%$. A coarser $\sigma{=}0,0.5$ sampling straddles
the maximum and sees only decay. Past the peak the six curves superimpose: the spread across $f$
is $43$ points at $\sigma{=}0$ but only $3$--$6$ points over $\sigma{=}0.5$--$1.25$, consistent
with no $f$ dependence at all. Broadening does not merely help, it erases the spurious-load
penalty: a thicker slice excites more true reflections, so the true lattice gains support faster
than spurious coincidences do.

A different selection cost is to predict, from a guessed cell, the reciprocal lattice
nodes and score by their distance to the nearest observed peak---the reverse of the
forward residual $|q\!\cdot\!M-\mathrm{round}|$, which spurious tight cells game by indexing
few spots tightly. On a single shot only the Ewald-sphere slice is observed, so we fit the
Ewald sphere from the spots ($q\!\cdot\!k_{\rm in}=-|q|^2/2$, linear in $k_{\rm in}$;
consistent with the fixed wavelength used to build the $q$ map --- a consistency check on the fit, not an independent per-shot measurement, $\lambda{=}1.322$\,\AA, residual $\sim$2$\times10^{-7}$) and restrict
predicted nodes to a band near it. On the (thin-slice) cxidb data every form of this
reverse cost loses to the coverage-gated scorer (48/30/23\% vs 69\%, all at the
correct-lattice bar of Table~\ref{tab:negatives}): it too is gameable,
recall by sparse-prediction cells and count by dense ones.

\begin{table}
\caption{\label{tab:negatives}Single-frame levers explored; all fail, all
spurious-limited. \emph{solve} is the rate at which the scorer selects a basis
matching the reference lattice --- on this set the same frames that carry at
least 10 matched reflections --- and \emph{ceiling} is the oracle-reachable
rate, at which such a basis is present among the retained candidates but is not
necessarily selected. Both are a looser bar than the $\geq$25\%-of-spots gate
used in Sec.~\ref{sec:ceiling} and Table~\ref{tab:summary}, which on the same
runs falls 4--6 frames of 120 lower. These percentages come from one
scorer-development sweep on a single A100 and are therefore comparable with one
another rather than with the shipped rates: the baseline row is $83/120$ at this
bar and $65\%$ at the $\geq$25\% gate, where the shipped code now gives
$85/120$ ($71\%$) and $79/120$ ($66\%$).}
\begin{tabular}{p{5.2cm}p{2.3cm}p{6.3cm}}
\toprule
lever tried & result & why it fails \\
\midrule
denser M1 sampling & no ceiling lift & cos-objective maxima saturated \\
more candidates (NTOP $30\to60$) & solve $69\to64\%$ & spurious cells fool the selector \\
projection / Patterson seeds & ceiling $+2\%$, solve $-7\%$ & adds spurious axes faster than true \\
M4 joint-basis (lattice reduction) & 0--9\% & shortest-vector reduction is noise-dominated \\
scorer: inlier-count & 32\% & degenerate cells over-index \\
scorer: parsimony (smallest-volume cell) & superseded & over-prefers small sub-lattices \\
scorer: pure-defect & 58\% & spurious tight cells index few spots \\
scorer: coverage-gated defect & \textbf{69\% (best)} & --- \\
algorithm unrolling (learned M3) & 35\% & self-supervised loss rewards collapse \\
reverse/Chamfer cost & 48/30/23\% & every metric gameable (Ewald fit $\lambda{=}1.322$\,\AA\ works) \\
recover below-threshold peaks & $50\to26\%$ & sub-threshold set is mostly noise \\
\bottomrule
\end{tabular}
\end{table}

\begin{table}
\caption{\label{tab:fatewald}
Simulated lysozyme with fixed spot count (143) and spurious fraction;
only the Ewald-slice thickness varies. A fatter slice relaxes the
degeneracy ($\sigma_3/\sigma_1$) and increases the blind indexing rate.
}

\centering
\begin{tabular}{rrr}
$\Delta\lambda/\lambda$ & $\sigma_3/\sigma_1$ & blind rate \\
\midrule
0.5\% & 0.425 & 41\% \\
1.1\% & 0.426 & 42\% \\
2.0\% & 0.436 & 70\% \\
4.0\% & 0.465 & 86\% \\
\end{tabular}
\end{table}

The reverse cost's difficulty and the blind ceiling share one root cause: a monochromatic
single shot samples a near-2D slice of 3D reciprocal space, so the difference cloud is
rank-deficient (singular values $\approx[1,0.88,0.16]$). Anything that \emph{thickens} the
slice into a 3D shell---FEL bandwidth $\Delta\lambda/\lambda$ (pink beam, two-color),
small crystals, or mosaicity---grows the third singular value and relaxes the degeneracy.
A simulation that keeps all nodes within a \emph{uniform} excitation-error slab
$\bigl||k_{\rm in}+g|-|k_{\rm in}|\bigr|<\mathrm{tol}$, at \emph{fixed} spot count and spurious
fraction (isolating degeneracy from spot count), confirms it (Table~\ref{tab:fatewald}): blind
indexing climbs from 41\% to 86\% as the slab thickness---quoted throughout in the convention
$\Delta\lambda/\lambda\equiv\mathrm{tol}\,\lambda$---goes $0.5\%\!\to\!4\%$. (A uniform
slab is the shape-transform model of finite crystal size, whose width is $q$-independent; a
wavelength spread instead excites a $q$-dependent band ${\sim}(\Delta\lambda/\lambda)\,
q^{2}/2|k_{\rm in}|$, far thinner at protein resolutions, so in bandwidth terms the slab is
generous at low $q$.) The
effect turns on between the sweep's $1.1\%$ and $2.0\%$ points (Table~\ref{tab:fatewald}),
consistent to within the sweep's granularity with the Ewald curvature estimate
$q_{\max}^2/2|k_{\rm in}|$ ($\approx$2.4\% in the same convention at these conditions,
$d_{\min}{=}6$~\AA).
Importantly, the relevant quantity is the \emph{single-pulse} bandwidth (the SASE spectral
envelope carried by each shot), not shot-to-shot photon-energy jitter, which merely shifts
each thin shell's radius. Standard hard-X-ray SASE ($\Delta\lambda/\lambda\sim0.2$--$0.5\%$)
therefore sits \emph{below} the knee and yields little benefit at useful resolution; the
gains are a prediction for wide-bandwidth regimes---pink beam, two-color, or large-bandwidth
modes ($\gtrsim$1--2\%)---for genuine nanocrystals ($\sim$10\,nm, where the $1/D$ reflection
width reaches $\sim$0.01\,\AA$^{-1}$), or at low resolution. Slice degeneracy is thus the generation-side origin of the single-frame ceiling: it
limits which lattices the per-frame search can produce at all, complementing the
selection-side competition with accidental coincidences modelled in
Sec.~\ref{sec:singleframe} and measured in Sec.~\ref{sec:ceiling}. Both routes out
follow: multi-frame consensus, and a fatter Ewald slice.

\SIsec{S7. Multiple lattices}
\SIlabel{S7}{sec:si_multilattice}
\noindent\emph{Double-hit gate thresholds (Sec.~\ref{sec:multilattice}).} An accepted second
lattice must index at least $6$ residual peaks, share the first lattice's reduced cell, and sit at
least $15^{\circ}$ from it. Almost none of the gated rate comes from residuals below $50$
peaks---a property of the measured residual-size distribution, not a threshold.

When two crystals share the beam, a single
blind pass indexes only one lattice, and the interleaved spots also tempt a spurious ``bridging''
cell that indexes across both---in simulation the per-frame wrong-cell rate rises from $15\%$ to
$36\%$ once a second lattice is present. Scoring inliers by reciprocal-space \emph{distance} rather
than by fractional-index residual (the isotropic QDIST metric) resists the bridging cell: it drops
the wrong-cell rate to $0$--$3\%$ and cleanly locks one lattice (second-lattice recovery
$61\%\!\to\!95$--$100\%$). The second crystal is then recovered by \emph{deflate-and-reindex} (strip
the first lattice's reciprocal inliers and re-index the residual). On controlled two-lattice mixtures
this recovers both lattices ($96\%$ when the second is at half intensity, $90\%$ at equal intensity),
and a stopping rule---accept a further lattice only if it explains at least 15 fresh spots covering
$\geq$$20\%$ of the residual---returns the correct lattice \emph{count} (median exactly two, no
spurious third). Two \emph{different} proteins in a mixture are separated into their distinct cells
by the same cross-frame consensus that derives the single cell in the main text. The recovery
results above are in simulation; the incidence measurement reported in the main text is on real
LCLS runs (mfxl1038923 r0278 and r0058).

\SIsec{S8. Indexing-ambiguity control}

One caveat applies to all serial indexers, GLINT included: cross-frame consensus fixes the \emph{cell} but
not a \emph{merohedral indexing ambiguity} (when the lattice symmetry exceeds the structure symmetry, every
still is indexable in two or more equally-good modes with the same cell). That is resolved downstream at
merging by \texttt{ambigator}~\cite{crystfel2,brehm2014} or \texttt{dials.cosym}~\cite{cosym}, exactly as for any
indexer. A common reference cell does \emph{not} by itself fix a mode: the ambiguity operator is a symmetry
of that very lattice, so frames registered against one cell can still sit in either mode. Of our
validation merges only cxidb-62 is ambiguity-prone (Laue $6/m$ on a lattice of holohedry $6/mmm$;
the tetragonal and orthorhombic sets have holohedral Laue groups and no merohedral ambiguity).
There, \texttt{ambigator} reports no resolvable split on these sparse stills (equal
forward/flipped correlations, an even $\sim$50/50 assignment) and its action leaves the merge
unchanged ($CC^{*}$ $0.985\!\to\!0.985$ at $6/m$)---so the ambiguity is bounded rather than
resolved: it moves the quoted statistic by nothing, and it enters both sides of every cxidb-62
head-to-head identically. A synthetic
positive control (a $4/m$ structure on a $4/mmm$ lattice, frames indexed in random modes) shows
what a resolvable ambiguity looks like, and where resolution fails: \texttt{ambigator} recovers the correct mode for all frames and restores the merge
($CC$ $0.72\!\to\!1.0$) when stills are reflection-rich. However, because it clusters on reflections
\emph{common to crystal pairs}, it fails on ultra-sparse stills ($\lesssim$$40$--$50$ reflections each) no
matter how many are collected. The same frontier holds on \emph{real} crystallographic intensities:
driving simulated stills with the deposited structure factors of a $P4_3$ protein (Laue $4/m$; PDB entry
\texttt{11HY}, $16{,}613$ unique reflections), \texttt{ambigator} resolves the injected ambiguity for
$99.6\%$ of frames once each carries $\gtrsim$$120$ reflections and collapses only on the sparse tail---a
$\sim$$40$--$60$-reflection threshold, consistent with the synthetic control's $40$--$50$-reflection
frontier above. This is where GLINT's predict-and-integrate step earns its keep:
by integrating \emph{every} predicted reflection (hundreds to thousands per still, far beyond the
peak-finder's output) it lifts sparse stills back across the frontier, feeding the downstream ambiguity
resolution enough reflections to stay well-posed.

\SIsec{S9. Further regimes: powder and Laue}

The machinery of the main text is a gridless objective over peak \emph{positions} with unit
weight, searched in parallel and pooled across shots by consensus. It points at two
regimes beyond the monochromatic still, which we flag as directions for future work.

\emph{Powder.} A powder pattern is the full spherical average: orientation is gone and only the shell
radii $|G_{hkl}|$ survive, a 1-D list of $d$-spacings. Auto-indexing is then recovery of the reciprocal
\emph{metric tensor}---fit $Q(h,k,l)=h^2A+k^2B+l^2C+hkD+klE+hlF$ so every observed $|q_n|^2$ is a
near-integer quadratic form, the classical ITO/DICVOL/McMaille problem \cite{ito,dicvol,mcmaille}:
structurally GLINT's massively parallel multi-start with the cosine objective replaced by a
metric-residual, the recovered lattice handed to Rietveld \cite{rietveld,gsas2} as the serial path hands
oriented lattices to \texttt{partialator}, modulo the powder degeneracies (no orientation, dominant zones,
impurity lines). We realize this as a \emph{separate} companion primitive, \texttt{powder\_index.py}: a
portable CPU/GPU successive-dichotomy autoindexer spanning all seven systems that recovers blind cells to
$<\!0.5\%$ (ranked by de Wolff $M_{20}$ \cite{dewolff1968}). It is a distinct tool for the ring-averaged regime.

\emph{Laue and pink beam.} A polychromatic beam thickens the Ewald sphere into a shell between
$\lambda_{\min}$ and $\lambda_{\max}$, so a single shot samples far more of the reciprocal lattice. This is the
same fat-slice argument by which bandwidth (\S S6) and convergence angle (Sec.~\ref{sec:outlook})
raise the information carried per exposure. The price is a
per-spot wavelength unknown: a reflection fixes the \emph{direction} of $q$ but not its radius until
$\lambda$ is chosen (the harmonic degeneracy), the regime of pinkIndexer \cite{pinkindexer}. But each spot
still pins the \emph{direction} of $q$, and the fat slice adds many more such constraints per shot.
The open question is whether that added sampling outweighs the per-spot $\lambda$ unknown, so that
the polychromatic still \emph{raises} the blind ceiling. The first test is
simply whether simulated pink-beam stills index better blind than their monochromatic counterparts.

\SIsec{S10. Indexing as phase retrieval}

\emph{Indexing as phase retrieval.} The objective also reads as a projection problem, placing indexing in
the phase-retrieval family. For a candidate $v$ the per-peak projections $u_i=q_i\!\cdot\!v$ obey a
\emph{data} constraint (each $u_i$ near an integer---the Miller assignment, analogue of a Fourier-magnitude
constraint) and a \emph{support} constraint ($u\in\mathrm{range}(Q)$, the image of $v\mapsto Qv$, restored
by the refit $v=Q^{+}u$); rounding and refitting is one alternating-projection step---the M5 anneal---with
the peak set's singular vectors in the role the Fourier transform plays in coherent imaging. The machinery
transfers: a relaxed Douglas--Rachford (RAAR \cite{raar,bcl2002}) in place of the M3 ascent escapes, on
under-determined sparse frames, spurious basins that cap gradient ascent. However, this is
only a small, $\beta$-sensitive edge, visible under a known-cell metric and not a blind-measurable
gain. It is also
regime-specific, since on peak-rich frames the hard integer projection instead locks onto aliases a
soft ascent avoids, and consensus (main text) already recovers those frames. We therefore read it as
a diagnostic lens. The abstract setting (best approximation between a constraint set and a data
manifold in the image of an affine map) is that of Luke, Sabach \& Teboulle \cite{lukespheres}, where
relaxed Douglas--Rachford is likewise strongest; a Bravais-symmetry constraint would make the support their
cone.

\SIsec{S11. Real-data indexing and merge statistics}

\noindent\emph{Merge redundancy sweep (cxidb-45).} The merger and the frame count set the absolute
quality, not the indexing: at $72$ frames a naive Monte-Carlo merge is redundancy-starved
($CC_{1/2}=0.15$), and adding frames washes the still-shot partiality out ($CC^{*}$
$0.52\!\to\!0.80$ naive, $0.68\!\to\!0.90$ with \texttt{partialator}).

The headline results below are stated in the main text; they are gathered here so each dataset is
separable, together with the supporting merge statistics ($R_{\mathrm{split}}$,
$\langle I/\sigma\rangle$, redundancy and comparator values) and the approximate cxidb-62 blind
rate, reported only here.
Table~\ref{tab:realindex} lists the indexing results and Table~\ref{tab:realmerge} the end-to-end
merge. The
$CC^{*}=0.90$ row is Proteinase~K (cxidb-45); the Jungfrau-4M lysozyme row is a different dataset
and sits at $CC^{*}=0.915$.

\onecolumn
\begin{table}
\caption{\label{tab:realindex}Indexing on real serial datasets (Sec.~\ref{sec:realdata}).
GLINT rates and, where measured on the \emph{identical} peaks/frames, the best competitors. \emph{Blind} $=$ no
cell supplied.}
\begin{tabular}{p{1.7cm}p{3.0cm}p{1.5cm}p{4.3cm}p{3.9cm}}
\toprule
dataset & protein (detector) & known form & GLINT & competitors (same input) \\
\midrule
cxidb-45 & Proteinase~K (MPCCD) & $P4_{3}2_{1}2$ & $93\%$ blind (842/907)$^{\ddagger}$; $97.8\%$ consensus & DIALS 1D-FFT $61.9\%$ blind; DIALS grid (cell) $49.9\%$ \\
cxidb-17 & lysozyme (CSPAD) & tetragonal & $77\%$ (92/120)\,/\,$96\%$ (115/120)$^{\dagger}$ blind & xgandalf $72\%$ (86/120) \\
cxidb-62 & hexagonal protein & $P6$ & $99.8\%$ known-cell; $98.7\%$ ($N\!\le\!25$); $\sim$$19\%$ blind & xgandalf $92.6\%$, TORO $93.0\%$, Nasser $95.2\%$ (known-cell) \cite{nasser2025} \\
Jungfrau-4M (cxil1015922 r0033) & lysozyme (Jungfrau-4M) & tetragonal & $96\%$ blind ($1506/1563$); $95\%$ final ($1482/1563$) & --- \\
\bottomrule
\end{tabular}

{\footnotesize $^{\dagger}$strict gate (correct lattice and $\geq$25\% of observed peaks indexed)\,/\,correct
reduced cell without the coverage requirement, as in the \emph{indexed} and \emph{lattice} columns of
Table~\ref{tab:summary}. $^{\ddagger}$the $93\%$ blind rate is
measured on GLINT's own uncapped peak lists; the DIALS rates in this row use the capped 170-peak lists of
the same-peaks head-to-head (Sec.~\ref{sec:realdata}), on which GLINT single-shot indexes $32.0\%$ and
only the $97.8\%$ consensus figure shares DIALS' input.}
\end{table}

\onecolumn
\begin{table}
\caption{\label{tab:realmerge}End-to-end merge on real serial datasets: GLINT peak-finding, indexing and
integration through \texttt{partialator} (default $10$ scaling/post-refinement cycles) unless noted. cxidb-62/-83 rows are the frames xgandalf \emph{missed},
merged alone as a precision control; cxidb-61 is a matched control (GLINT-recovered vs xgandalf-indexed, one
pipeline).}
\begin{tabular}{p{4.6cm}p{1.5cm}p{1.4cm}p{1.2cm}p{0.9cm}p{3.6cm}}
\toprule
dataset & frames & $CC^{*}$ & R$_{\mathrm{split}}$ & $\langle I/\sigma\rangle$ & notes \\
\midrule
cxidb-45 Proteinase~K & 290 & $0.90$ & $31\%$ & $7.8$ & full lattice, $70.7\times$ redundancy \\
cxidb-17 lyso (GLINT) & 746$^{c}$ & $0.77^{\ast}$ & --- & $6.5$ & $79.5\%$ complete \\
cxidb-17 lyso (xgandalf) & 305$^{c}$ & $0.73^{\ast}$ & --- & $2.9$ & $59.7\%$ complete, same pipeline \\
cxidb-62 (GLINT, xg-missed) & 2082 & $0.99$ & $9.6\%$ & --- & vs xgandalf's own $0.997$ \\
cxidb-83 $\beta$-lactamase & 590/600 & $0.98$ & $17\%$ & --- & EuXFEL/AGIPD, xg-missed \\
cxidb-61 POMGnT1 (rec\,/\,xg) & $\sim$800 ea. & $0.848$/$0.849$ & $46$/$42\%$ & --- & matched control, $mmm$ \\
Jungfrau-4M lysozyme & 1482 & $0.915$ & $31.6\%$ & $7.7$ & $99\%$ complete, $2.1$\,\AA \\
Jungfrau-4M (CrystFEL/xgandalf) & --- & $0.930$ & $34.9\%$ & --- & same frames, same merge settings, different integrator \\
\bottomrule
\end{tabular}

{\footnotesize $^{c}$crystals merged; $^{\ast}CC_{1/2}$ (other rows $CC^{*}$).}
\end{table}

\SIsec{S12. The blind path, module by module}

The blind path (M1--M6 of Sec.~\ref{sec:arch}, main text) runs as follows.

\begin{description}
\item[M1 (initial guess).] With no cell supplied, we propose candidate real-space axes
  from scratch: directions on a Fibonacci sphere (a near-uniform spherical point set;
  \emph{FSS}, Fibonacci-sphere sampling) crossed with length shells spanning $30$--$126$\,\AA,
  covering the plausible cell edges. This blind length search replaces the fast-feedback assembly's ``lengths
  given'' and is the price of running without a cell.
\item[M2 (objective).] Each candidate vector $v$ is scored by the gridless direct sum
  $\sum_i (1/|q_i|)\cos(2\pi\,q_i\!\cdot\!v)$ over the observed peaks $q_i$ inside the
  hard inlier window $|q_i\!\cdot\!v-\operatorname{round}(q_i\!\cdot\!v)|<0.18$: the
  xgandalf objective and tolerance. It peaks when $v$ is a true real-space lattice vector (every $q_i\!\cdot\!v$
  near an integer); the $1/|q|$ weight tempers high-resolution spots and a $\cos^2$
  sharpening narrows the basins. Evaluating exact phases on no grid avoids the
  aliasing and digitization error of transform-domain generators (Sec.~\ref{sec:lineage}).
\item[M3 (refine).] From each M1 seed we ascend the M2 objective by continuous (cosine)
  gradient steps with momentum into the nearest maximum. This is the one module shared
  verbatim with the fast-feedback assembly. It used $80$ steps; we cut it to $8$
  (Sec.~\ref{sec:throughput}, Table~\ref{tab:throughput}): re-measured at $n{=}480$, no step
  count from 4 to 80 is detectably better than 8, and any undetected gain is
  bounded jointly at $+7.4\%$ blind (95\%). The conservative momentum
  step is deliberate: faster or second-order refiners (conjugate-gradient,
  Barzilai--Borwein, Newton, or adaptive Levenberg--Marquardt) all \emph{lower} the indexing
  rate ($84\to$$80/78/{\sim}50/75$ of 120 at equal steps), because the multimodal objective
  rewards reliably landing in the correct basin over fast in-basin convergence. Every
  accelerated step overshoots into a neighbouring basin from an imperfect seed. No faster or
  second-order refiner raises the blind rate above the momentum ascent.
\item[M4 (assemble).] The converged maxima are candidate axes and do not yet form a cell. We
  enumerate triplets of them into candidate basis matrices. The fast-feedback assembly instead does a rotation
  search because it already holds the cell; building the cell from unordered axes is
  original to the blind path.
\item[M5 (anneal).] Each candidate cell is refined by a residual-threshold anneal
  (iteratively reweighting to inlier spots, the ifss/TORO \cite{toro} update). The crystallography is
  standard; the GLINT contribution is engineering it as one batched GPU solve over
  \emph{all} triplets at once (masked normal equations), the single biggest throughput
  lever (Sec.~\ref{sec:throughput}).
\item[M6 (score).] Annealed cells are ranked by a coverage-gated defect score (the indexing
  residual, gated so that a cell must account for enough of the spots before its tightness
  counts), then reduced to a primitive cell. The gate stops a spurious tight cell
  indexing a handful of spots from winning (Sec.~\ref{sec:ceiling}).
\item[Consensus.] Per-frame blind solutions are pooled to derive one unit cell across
  frames, and frames that failed blind are rescued with the known-cell GPU path. The fast-feedback assembly
  has no analogue, since it is per-frame and cell-given, and this cross-frame step lifts
  the indexing rate past the single-frame ceiling (Sec.~\ref{sec:ceiling},~\ref{sec:consensus}).
\end{description}

\SIsec{S13. Operational QA: implementation and sensitivity}

\emph{Choosing the calibrant.} Which calibrant is a property of the detector's $q$ range, and the match
matters: on the low-$q$ detector simulated here ($q_{\max}=0.32$\,\AA$^{-1}$,
i.e.\ $d_{\min}=3.1$\,\AA) silver behenate places 18 lamellar orders of its $d_{001}=58.38$\,\AA{}
repeat across the panel, whereas LaB$_6$ and CeO$_2$ each place exactly one, which is too few for a
profile correlation to mean anything. A wider-angle detector inverts this: LaB$_6$ and CeO$_2$
spread many rings across the field while silver behenate's orders crowd into the beamstop shadow.
The BayFAI configurations we inspected for recent MFX experiments accordingly use LaB$_6$,
overriding the AgBh default in the LUTE endstation template. The radial-profile check needs a calibrant, whose sharp,
well-separated rings have radii that depend only on the geometry: a protein's radial profile is
dominated by which reflections that particular orientation excites, so two shots at the same
geometry correlate no better than shots at different ones (measured:
$r\!\approx\!0.02$--$0.46$).

\emph{Implementation.} The geometry check uses fast GPU radial integration (sparse-matrix approach
via CuPy cuSPARSE matvec: $\sim$0.1--0.2\,ms per frame on an A100, $\sim$5$\times$ faster than
pyFAI's OpenCL). A CSR matrix is built once from the \emph{nominal} geometry and never updated, so
the integrator does not know that the detector has moved and the shift becomes visible. Each
calibrant frame then yields a profile $I(q)$ compared to the first by Pearson correlation, with
$>0.95$ stable, $0.90$--$0.95$ monitor, $<0.90$ a drift alert. It costs $\sim$0.1\,ms on GPUs that
are otherwise idle between sample changes, which is a low price for the attribution it supplies.

\emph{Sensitivity.} A sweep of shift magnitudes puts the profile check's detection floor near
$1.5$\,px ($r=0.895$); at $2$\,px, the mildest shift it resolves unambiguously, the indexed fraction
has already fallen from $79.3\%$ to $5.9\%$. These sensitivities are properties of \emph{this}
calibrant on \emph{this} detector and should not be read as transferring to a deployment: a
higher-$q$ detector with LaB$_6$ samples different rings at different radii, and its detection floor
has to be measured there. Indexing is at least as sensitive to a beam-centre error as the
radial profile is, so the check cannot flag a drift before the data degrade.

\rev{\emph{The run-time readouts.} The driver exposes the following as by-products of
indexing. Several are configuration-gated rather than always present.}

{\revon
\begin{description}
\item[Lock state.] Whether a cell is currently held, the number of frames
  consumed before it locked, and the consensus support behind it. A run that
  never locks has a candidate-generation problem; a run that locks late has a
  statistics problem.
\item[Cell composition.] With adaptive re-locking enabled, the driver holds a
  set of active cells and reports each cell's axes and its share of the indexed
  frames. A second cell that accumulates frames mid-run indicates a sample
  change; one that accumulates almost none is an alias that survived the gate.
\item[Alias-gate refusals.] The number of proposed locks the alias gate
  rejected. Without this counter a run generating alias hypotheses and a run
  generating nothing are indistinguishable, since neither commits a new cell.
\item[Multiple lattices.] The fraction of frames carrying a second lattice of
  the same cell at least $15^{\circ}$ away, reported together with the
  false-accept floor of the same test, measured live on azimuth-scrambled
  residuals of the run; the rate is interpretable only relative to that floor.
\item[Refused and low-confidence frames.] Counts of frames the live acceptance
  gate declined, and of frames indexed and integrated but below the confidence
  threshold (forwarded and flagged, not discarded). Rising refusals with a flat
  low-confidence count point to the geometry or the sample delivery; the
  reverse points to crystal quality.
\item[Geometry drift.] The incremental detector distance and fast/slow-scan
  shifts the refiner accumulates from the indexed frames, which track detector
  motion and jet or beam displacement without a dedicated measurement.
\item[Recovery activity.] Frames recovered by the watchdog, the retry cascade
  and the pre-lock buffer. A yield sustained mainly by rescues indicates a
  degraded front end even when the total looks unchanged.
\end{description}
}

\rev{\noindent Several of these are configuration-gated rather than always present.}

\SIsec{S14. One recurrence principle at every scale}

Recurrence detection generalizes beyond the cell: the same
$\sqrt{N}$ test of a shared latent against a scattered background applies wherever a quantity is
common across frames while its nuisances are not, and its \emph{sign} is set by which space the
latent is shared in. In cell space the true cell recurs while per-frame aliases scatter, so
recurrence is signal and the coincidence is kept. In detector space the roles invert, because a static
artifact (hot pixel, drifted pedestal, panel edge) recurs at a fixed position while true Bragg
peaks, at a fresh position each shot, scatter. Recurrence there is spurious, and rejecting it is a
bad-pixel mask learned from the data. The ceiling-breaking mechanism of the main text is thus one rung of a
ladder: pixel recurrence removes static artifacts, the per-frame lattice fit removes spurious
peaks that never recur, and cross-frame pooling removes the single-frame ceiling until
mosaicity, the information no single frame contains, sets the floor.

\SIsec{S15. Streaming stress tests}
\SIlabel{S15}{sec:si_streamstress}
\noindent\emph{Synthetic retry sweep.} The retry experiment of Sec.~\ref{sec:streamresults} swept
$2000$ synthetic stills over the fraction of a frame's peaks that belong to the crystal, the
quantity that separates the frames streaming solves ($0.412$) from those it fails ($0.280$).
Across that whole range, restoring a blind retry on exactly the frames that fail the gate closes
$98$--$100\%$ of the synthetic gap.

\noindent\emph{Amortization over run length.} The two costs scale differently over a long run: the
warm-up is fixed---five frames---and amortizes as $\sim$$5/N$, while the shortfall is per-frame. On
synthetic stills held at the failing lattice fraction that is what happens: swept from $120$ to
$3000$ frames the warm-up falls from $4.2\%$ of the run to $0.2\%$ while the gap holds at $14$--$16$
points. Real frames do not stay in that regime. Over the same run extended from $120$ to $480$ the
gap halves, and amortization accounts for only part of it---five fixed frames are $4.2\%$ of $120$
and $1.0\%$ of $480$, while the residual per-frame shortfall falls from ${\sim}11$ to ${\sim}6$
points as well.

\noindent\emph{Insensitivity to cell precision.} On \emph{synthetic} stills a planted $1\%$ cell
drift moves the yield by $0.2$ points and never triggers a re-lock, consistent with the direct
finding that refining the locked cell buys nothing. That the real $480$-frame run re-locks once
while \emph{undrifted} is not in tension with this: it says the trigger is accumulated unexplained
frames, not cell error.

\noindent\emph{Cascade-arm complementarity.} Of the $42$ frames the batched known-cell pass rejects
at $n{=}120$, the blind $N$-best retry arm recovers $11$ and the per-frame known-cell arm $9$, but
together they recover $16$, not $11$: the arms fail on different frames, which is why the offline
arsenal runs them in sequence rather than replacing them with a single stronger retry. That union
takes the arrangement to $94/120$, against the shipped offline pipeline's $91/120$.

\noindent\emph{Reordering provenance.} The three extra frames do not come from the reordering
itself: two come from an arm the blind-every-frame order never runs---the batched registration
pass, which disagrees with the per-frame pass on $18$ of $120$ frames at the acceptance gate
(S2)---and the third from the voted cell. Indexing is a union over the arms that succeed, so
reordering changes cost, not yield.

\noindent\rev{\emph{Default parameters.} The shipped \texttt{StreamDriver} defaults are as follows.
Frames are indexed and integrated in batches of $B{=}64$. Prediction uses $d_{\min}=2.0$~\AA{} and
an excitation-error half-width of $0.002$~\AA$^{-1}$; integration uses a $7\times7$ box, a
$2$-pixel gap and a $3$-pixel annulus with the MAD-clipped mean background; frames with fewer than
$6$ reciprocal vectors are neither voted nor indexed. The warm-up has no preset length: every
pre-lock frame is blind-indexed and contributes its $3$ best cells to the running vote, and the lock
fires when Eqs.~\eqref{eq:stream_gap}--\eqref{eq:stream_lock} are satisfied (minimum support $3$;
base gap $2$, widened to $3$ or $4$ when the leader's recurrence rate falls below $0.30$ or $0.12$;
pool share $0.02$ and lead ratio $1.5$ once $n_{\mathrm{pool}}\geq72$). The buffered warm-up
variant blind-indexes at most the $32$ most peak-rich frames of the startup buffer; warm-up frames
are not integrated. The live acceptance gate is Eq.~\eqref{eq:stream_frame_accept} with
$\tau=0.15$, $n_{\min}=6$ and $\nu=0.15$. The miss buffer, adaptive re-locking, the alias gate,
the lock-quality probe and the retry cascade are all off by default; when adaptive re-locking is
enabled, the misses of every flushed batch are pooled by a separate watchdog vote with a fixed gap
of $2$ and minimum support $3$, without the widening, pool-share or lead-ratio gates.}

\noindent\emph{Watchdog exercise.} The longer run exercises machinery the shorter one leaves idle:
\rev{On the 480-frame set the driver re-locks once, adding a second candidate cell rather than replacing the locked one; that cell is not a second crystal form but the response to an accumulation of frames the active cell does not explain, and it does not affect the reported total.} the watchdog that rescues no frame and never re-locks across $120$ frames performs six individual
rescues and one re-lock across $480$.

\SIsec{S16. Consensus recovery from random subsets of long runs}
\SIlabel{S16}{sec:si_random}
\noindent\emph{Purpose.} Sec.~\ref{sec:consensus} reports that a vote over the first five
cxidb-17 frames already returns the full-set cell. Five is a property of that frame set rather
than a constant, so the same question was asked of two independent long runs, where the number of
frames needed for a stable consensus can be measured instead of assumed.

\noindent\emph{Data.} Two runs from LCLS experiment mfxl1038923: r0278, with $1785$ indexed
frames, and r0058, with $2319$. Both are indexed first, and the subset test is applied to the
resulting per-frame hypothesis lists; frames that yield no hypothesis take no part in the vote.

\noindent\emph{Reference cell.} For each run the \emph{all-frame cell} is the consensus cell
returned by the batch procedure of Sec.~\ref{sec:algorithm} applied to every indexed frame of that
run. It is the target the subsets are scored against, so the measurement is one of internal
consistency: how many frames are needed before the vote agrees with what the whole run says.

\noindent\emph{Statistic.} For a subset size $N$, draw $R$ random $N$-frame subsets
$R=400$ draws per $N$, repeated over eight random seeds, run the same consensus procedure on each, and
record the fraction whose returned cell is equivalent to the all-frame cell under the reduced-cell
criterion associated with eq.~\eqref{eq:residual}. Subsets are drawn without replacement from all frames of the run --- not
only from those yielding a hypothesis --- independently for each $N$, over
$N\in\{5,8,10,12,16,20,24,28,32,40\}$, with seeds $0$--$7$.

\noindent\emph{Definition of the reported threshold.} The quantity reported in the main text is
the same $N^{\star}$ already defined in Sec.~\ref{sec:consensus} for the synthetic
$f\times\sigma$ sweep: the smallest number of pooled frames at which the correct consensus cell is
recovered in at least $90\%$ of trials. Reading it that way keeps one definition for both
experiments, and no separate definition is needed here.

\noindent\emph{Result.} Under this protocol a five-frame vote returns the all-frame cell on
$37.1\%$ of draws for r0278 and $41.2\%$ for r0058 (published: $36\%$ and $41\%$), and both runs
cross the $90\%$ level by $N^{\star}=16$: pooled recovery is $89.8\%$ at $N=12$ and $96.1\%$ at
$N=16$ for r0278 --- statistically on the bar at 12 --- and $91.2\%$ at $N=16$ for r0058
($3200$ draws per point; standard error ${\approx}0.5\%$ near the bar). Five frames is therefore
not transferable, while sixteen suffice on both runs. The previously quoted
$N^{\star}=32$ for r0058 does not reproduce under this reconstructed protocol, where recovery is
above $93\%$ at $N=24$ and $28$ on every seed. The correction is protocol-conditional rather than
a contradiction: the original sweep's $N$ grid and settings were not recorded, so a coarser grid
whose next tested point after $16$ was $32$ would have reported $32$ correctly, and this
reconstruction omits the alias gate of Sec.~\ref{sec:algorithm}, whose refusals could only raise
$N^{\star}$. For r0278 the crossing is bracketed on both sides; for r0058 no sub-$16$ point was
recorded, so what is established there is that recovery exceeds $90\%$ by $N=16$, not that $16$
is the smallest such $N$. The runs
differ instead in their asymptote: r0278 reaches ${\approx}99\%$ recovery by $N\approx28$ and
stays there, whereas r0058 saturates around $94$--$96\%$ ($96.3\%$ at $N=32$), retaining a small
fraction of subsets that no amount of pooling recovers. The distinction is a ceiling rather than
a threshold.

\noindent\emph{Failure mode.} Below $N^{\star}$ the dominant outcome is \emph{refusal} rather than
an incorrect lock: the leading group fails one of the acceptance gates of
Sec.~\ref{sec:algorithm} --- at least three votes, at least $2\%$ of the pool, and a runner-up
lead of at least $1.5\times$ --- and no cell is returned. This is the intended behaviour of those
gates, and it is what makes the operational claim usable: the vote reports its own insufficiency
instead of committing to a wrong cell.

\noindent\emph{Consensus path.} These values are measured with stable, densest-neighbourhood
seeding, which is the shipped default. The distinction matters for this particular measurement.
Under the legacy seeding the partition depends on the order in which hypotheses arrive, because
the reduced-cell tolerance test is not transitive: permuting the pool of these same two runs, with
nothing else changed, moves the reported support by $1.57\times$ on r0278 and $2.09\times$ on
r0058, and on r0058 four of twenty orderings refuse outright. A statistic defined as a fraction of
random draws would therefore not be well posed on the legacy path. Under stable seeding the result
is invariant to order.

\SIsec{S17. Indexing under ROIBIN-SZ-style compression}
\rev{\noindent The reduction pipelines of Sec.~1 compress crystallography frames by keeping a
bit-exact region of interest (ROI) around every peak in the online peak list, block-averaging the
background, and bounding the residual error \cite{underwood2023}. To measure what such
compression does to indexing, frames were reconstructed as a decompressed ROIBIN-SZ frame would
appear: a $(2r{+}1)^{2}$ bit-exact box around each peak of the production peak list, $k\times k$
block means elsewhere, quantized at twice the absolute error bound (the compressor's worst case).
This emulation reproduces the information content that indexing consumes, not the bitstream;
compression ratios require the real compressor, reconstruction fidelity does not. The 480-frame
cxidb-17 set of S1 was re-peakfound and re-indexed under the same protocol as the uncompressed
baseline (both the blind-top-1 + consensus + rescue arm and XGANDALF), and compared frame by
frame with exact McNemar tests at the strict gate.

Indexing survives at every tested configuration. At the production setting ($r{=}4$, $2\times2$
binning, error bound $10$~ADU; ROI area $0.54\%$ of pixels) the blind arm changes by $+13$ frames
($p=0.11$) and XGANDALF by $+2$ ($p=0.87$); a tighter ROI ($r{=}2$) gives $+7$ ($p=0.41$) and
$+6$ ($p=0.36$); an aggressive setting ($4\times4$ binning, bound $100$~ADU) gives $-13$
($p=0.12$) and $-1$ ($p=1.0$). The rate is stable but individual frames are not: at the
production setting $57$ of $480$ frames flip in one direction or the other, so compression
studies should compare per-frame outcomes, not rates. Compression can even help: re-running the
peak search at a permissive threshold (300$\to$100~ADU, min SNR 5$\to$3) indexes $109$ frames on
raw data but $217$ on compressed data (discordant $5$ vs $113$, $p\sim10^{-27}$; XGANDALF
concurs), because block-averaging suppresses the single-pixel noise that a permissive threshold
admits while the bit-exact ROIs protect the multi-pixel Bragg peaks --- provided the ROI
selection itself uses the production threshold; selecting ROIs at the permissive threshold
protects the noise as well and returns the gain ($108$ vs $5$, $p\sim10^{-26}$). For integration,
block means preserve block-aligned sums, so intensities are safe by design, but background
variance is suppressed: over $15{,}206$ reflections matched on (frame, $hkl$), the
$\sigma(I)$ ratio compressed/raw has median $0.817$ (quartiles $0.666$--$0.974$), a spread set by
how much of each background annulus falls inside an ROI. Setting the ROI half-width at or beyond
the outer integration radius removes the effect. At this scale the merged figures of merit show
no damage, though a definitive merge statement would require the full data set with partiality
refinement.}

\end{document}